\documentclass[a4paper,fleqn]{cas-dc}

\usepackage[numbers,sort&compress]{natbib}
\PassOptionsToPackage{no-math}{fontspec}
\usepackage{fontspec}
\usepackage{amsthm}
\usepackage{algorithm}
\usepackage{algpseudocode}
\usepackage{placeins}
\usepackage{rotating}
\usepackage{adjustbox}
\usepackage{float}
\usepackage{capt-of}
\usepackage{flushend} 

\usepackage{tikz}
\usetikzlibrary{positioning,arrows.meta,fit,calc,backgrounds,decorations.pathreplacing,patterns}
\usepackage{pgfplots}
\usepackage{pifont}
\pgfplotsset{compat=1.18}
\usepgfplotslibrary{groupplots}

\usepackage{etoolbox}
\newcommand{\figtablefonts}{
  \setmainfont{texgyretermes-regular.otf}[BoldFont=texgyretermes-bold.otf, ItalicFont=texgyretermes-italic.otf, BoldItalicFont=texgyretermes-bolditalic.otf]%
  \setsansfont{texgyretermes-regular.otf}[BoldFont=texgyretermes-bold.otf, ItalicFont=texgyretermes-italic.otf, BoldItalicFont=texgyretermes-bolditalic.otf]}
\AtBeginEnvironment{figure}{\figtablefonts}
\AtBeginEnvironment{figure*}{\figtablefonts}
\AtBeginEnvironment{table}{\figtablefonts\footnotesize\setlength{\tabcolsep}{4pt}}
\AtBeginEnvironment{table*}{\figtablefonts\footnotesize\setlength{\tabcolsep}{4pt}}
\AtBeginEnvironment{algorithm}{\figtablefonts}

\ExplSyntaxOn
\NewDocumentCommand \castablestyle {} { \__reset_tbl: \dim_set:Nn \l_tbl_width_dim { \linewidth } \cs_set_eq:cN { @makecaption } \__make_tbl_caption:nn }
\ExplSyntaxOff
\makeatletter
\renewcommand\paragraph{\@startsection{paragraph}{4}{\parindent}%
    {10pt \@plus0.01pt \@minus0.01pt}%
    {-1em}%
    {\ssssparaindent\ssssectionfont\upshape\bfseries\raggedright}}
\makeatother

\floatname{algorithm}{Algorithm}
\newcommand{\sraw}{s_{\mathrm{raw}}}

\ExplSyntaxOn
\cs_set:Npn \__first_footerline: { }
\exp_args:Nc \patchcmd { environment~keywords~code }
  { A\,R\,T\,I\,C\,L\,E\ \ I\,N\,F\,O \par \skip_vertical:n { -3pt } }
  { } { }
  { \PackageError{cfvul}{Could~not~remove~the~CAS~article-information~heading}{} }
\ExplSyntaxOff
\hypersetup{pdfcreator={LaTeX}}

\newcommand{\sstd}{s_{\mathrm{std}}}
\newcommand{\Kref}{K_{\mathrm{ref}}}
\newcommand{\Laux}{\mathcal{L}_{\mathrm{aux}}}
\newcommand{\Lbce}{\mathcal{L}_{\mathrm{BCE}}}

\begin{document}
\let\WriteBookmarks\relax
\def\floatpagepagefraction{1}
\def\textpagefraction{.001}

\shorttitle{Mitigating fix reversal with counterfactual constraints}
\shortauthors{Bian et al.}

\title[mode = title]{Mitigating fix reversal: Counterfactual constraints for function-level vulnerability detection}

\author[1,2]{Yun Bian}
\author[1,2]{Shihao Li}
\author[1,2]{Haiquan Wang}
\author[1,2]{Yi Chen}
\author[1,2]{Zhao Yang}
\author[1,2]{Zhe Cui}

\affiliation[1]{organization={Chengdu Institute of Computer Applications, Chinese Academy of Sciences},
            city={Chengdu},
            country={China}}
\affiliation[2]{organization={University of Chinese Academy of Sciences},
            city={Beijing},
            country={China}}

\begin{abstract}
Function-level vulnerability detectors often exhibit fix reversal, assigning a higher vulnerability probability to a fixed version than to its vulnerable counterpart. Fixes usually lengthen functions, and larger length increases are associated with more severe reversal. Code-transformation experiments indicate that fix reversal mainly stems from a length shortcut: vulnerability probability increases with function length when content is fixed. We propose CFVul to address this problem. Sliding-window encoding covers complete functions, retaining fix information beyond the truncation boundary. During training, a content constraint encourages the vulnerable version to receive a higher vulnerability probability under a common length input. During inference, a length constraint averages vulnerability probabilities over fixed reference lengths, reducing the influence of actual length changes. On PrimeVul, CFVul achieves a fix-pair ranking metric, PairRank, of 59.4\%, exceeding the existing methods compared. For fixes that increase function length by more than 15\%, PairRank rises from CodeBERT's 25.8\% to 48.4\%. PairRank also improves in experiments with four additional models and three additional datasets. By weakening the length shortcut, CFVul reduces cases in which fixed versions receive higher vulnerability probabilities than vulnerable versions. It requires only existing fix pairs and applies to multiple pretrained code models, facilitating adoption in existing detectors.
\end{abstract}


\maketitle

\section{Introduction}\label{sec:intro}

Exploited software vulnerabilities can expose data or compromise system control, making early vulnerability detection in source code essential to software security.
Machine learning for vulnerability detection (ML4VD) automatically identifies potentially vulnerable program units from source code. This allows limited code-review and testing resources to prioritise high-risk code~\citep{zhou2019devign,chakraborty2022deep}.
Deep-learning approaches typically fine-tune detectors based on pretrained code models~\citep{ding2025primevul}. They formulate vulnerability detection as function-level binary classification: each training function carries a binary label, and the detector predicts whether an input function contains a vulnerability. Its output is the estimated probability that a function is vulnerable, hereafter termed vulnerability probability. Performance is evaluated using conventional classification metrics such as F1 and the area under the receiver operating characteristic curve (AUROC).

Prior studies show that detectors can achieve high conventional classification metrics by exploiting spurious correlations in datasets. These metrics alone therefore provide insufficient evidence that detectors recognise vulnerabilities~\citep{chakraborty2022deep,risse2025top}.
Examining the basis of a detector's decisions requires appropriate comparisons between inputs.
In causal inference, a counterfactual is the outcome for the same individual under an alternative intervention~\citep{pearl2009causality}.
In vulnerability detection, a fix constitutes such an intervention. A detector that bases its decisions on vulnerabilities should assign a lower vulnerability probability after a function is fixed.

\begin{figure*}[width=\textwidth,pos=!t]
\centering
\includegraphics[width=\textwidth]{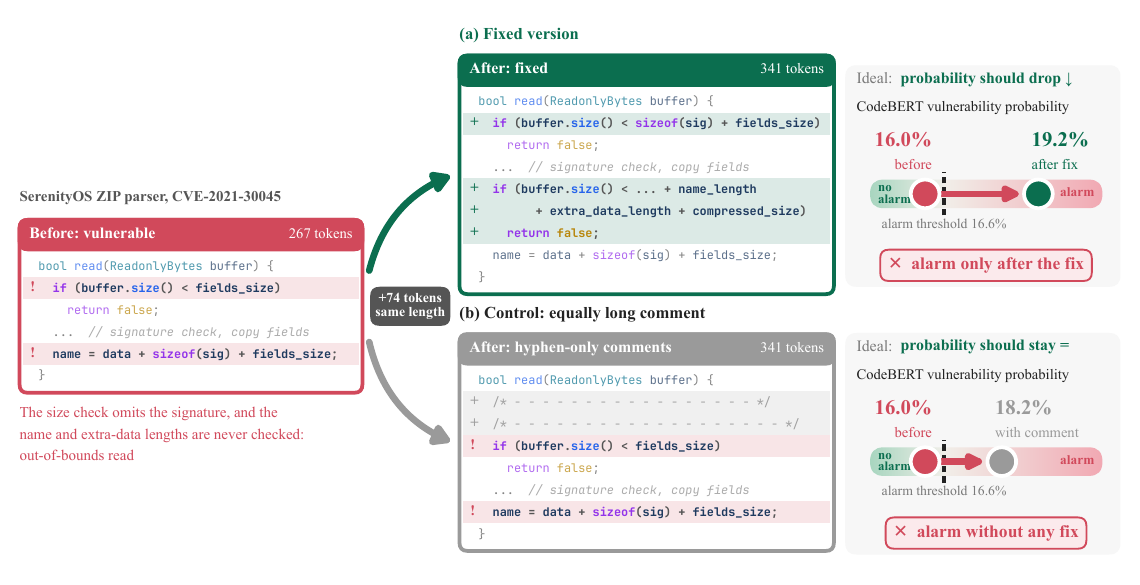}
\caption{Fix reversal in SerenityOS commit 4317db7 from the PrimeVul validation set; code is abridged. (a) Fixed version. (b) A control that inserts comments at the fix location in the vulnerable version, matching the length of the added fix code. The right-hand panels show CodeBERT's vulnerability probabilities. Dashed lines mark the alarm threshold that maximises validation F1.}
\label{fig:teaser}
\end{figure*}

Function-level datasets such as PrimeVul retain both versions of the same function before and after a fix. These versions form a fix pair, allowing direct comparison of the detector's outputs.
This raises the question of whether existing detectors assign higher vulnerability probabilities to vulnerable versions within fix pairs.
Figure~\ref{fig:teaser}(a) shows a fix pair from SerenityOS's ZIP parsing function. The vulnerable version omits the file header signature when checking buffer length and does not check the filename or extra-field lengths. These omissions can cause an out-of-bounds read. The fix includes the signature in the buffer-length check and adds checks for the filename and extra-field lengths. Although the fixed function should be safer, fine-tuned CodeBERT assigns it a higher vulnerability probability that exceeds the alarm threshold. The vulnerable version raises no alarm, whereas the fixed version does.
We call the assignment of a higher vulnerability probability to a fixed version than to its vulnerable counterpart fix reversal. We measure fix reversal at dataset level using PairRank, the percentage of fix pairs in which the vulnerable version receives a higher vulnerability probability. Lower PairRank indicates more severe fix reversal.
This behaviour also occurs across the evaluated detectors and fix pairs.
On PrimeVul's test fix pairs, CodeBERT, UniXcoder, GraphCodeBERT and CodeT5 all have PairRank below 50\%, indicating severe fix reversal. PairRank decreases as the length increase caused by a fix grows. When fixes increase function length by more than 15\%, CodeBERT assigns the vulnerable version a higher vulnerability probability in only 25.8\% of pairs.

Besides changing vulnerability-related code, fixes usually increase the length of a function.
Figure~\ref{fig:teaser}(b) inserts word-free comments at the fix location, using the same token count as the added fix code. The function remains unfixed, yet CodeBERT's output probability also increases and exceeds the alarm threshold.
This observation suggests a length shortcut: with content fixed, the detector's vulnerability probability increases with function length.
We designed two groups of code-transformation experiments to distinguish length changes from the fix itself. Length padding progressively inserts semantics-preserving comments at the fix location in the vulnerable version until it becomes longer than the fixed version. This increases length while leaving the underlying program content unchanged. The equal-length comparison pads the vulnerable version to the length of the fixed version, allowing comparison of the fix content at equal length. The results show that padding alone can make detectors assign higher vulnerability probabilities to vulnerable versions. At equal length, some detectors still struggle to distinguish versions based on fix content. Fixes increase function length, and the length shortcut raises the vulnerability probability of the fixed version. This makes the shortcut a major source of fix reversal that conventional evaluation does not readily quantify. In PrimeVul, vulnerability prevalence increases monotonically with function length. A length-only ranking rule achieves AUROC comparable to fine-tuned CodeBERT. Assigning higher vulnerability probabilities on the basis of length therefore need not noticeably reduce overall metrics. We introduce PairRank to measure fix reversal across datasets and expose length shortcuts obscured by conventional classification metrics.

Existing approaches that use fix pairs, such as SVulD~\citep{ni2023svuld}, separate the representations of the two versions through contrastive learning. They do not specify which version should receive a higher vulnerability probability. They also exhibit severe fix reversal on PrimeVul's test fix pairs.
To address fix reversal arising mainly from the length shortcut, we propose Counterfactual Constraints for Vulnerability Detection (CFVul) for function-level vulnerability detection.
Pretrained code models such as CodeBERT read only the first 510 tokens of a function. When a long function's fix lies beyond this truncation boundary, the two versions provide identical content to the model.
CFVul uses sliding windows to cover complete functions, exposing fix code beyond the prefix truncation boundary. Its classification head receives both a code representation and an explicit length input.
During training, the content constraint assigns both versions of a function a common length input. A pairwise loss encourages higher vulnerability probability for the vulnerable version. During inference, the length constraint substitutes fixed reference lengths for the actual length and averages the resulting vulnerability probabilities. Actual length therefore no longer affects the inference output through the explicit length input.
At inference, CFVul requires only a single function, without code-change information or a corresponding fixed version.
The primary evaluation metrics are PairRank and length-matched AUROC. They assess the ordering of the two versions and discrimination between functions of similar length, respectively. Conventional classification metrics support comparison with prior work.

We investigate the following four research questions in the experiments reported below:
\begin{enumerate}
\item[RQ1] \textbf{Is severe fix reversal widespread among existing detectors, and what causes it?}
\item[RQ2] \textbf{Can CFVul weaken the length shortcut and mitigate fix reversal?}
\item[RQ3] \textbf{What role does each component of CFVul play?}
\item[RQ4] \textbf{Can CFVul be applied to other encoders and datasets, and under what data conditions does fix reversal become severe?}
\end{enumerate}

The main contributions of this study are as follows:
\begin{itemize}
\item We identify and analyse fix reversal in function-level vulnerability detection. Detectors assign higher vulnerability probabilities to fixed versions, and reversal becomes more severe as fixes increase function length further. Length padding and equal-length comparisons indicate that fix reversal mainly stems from a length shortcut. Length alone achieves AUROC comparable to fine-tuned CodeBERT, obscuring the shortcut under conventional classification metrics.
\item We propose CFVul for function-level vulnerability detection using counterfactual constraints. Sliding windows cover complete functions, and the classification head receives an explicit length input. During training, the content constraint encourages higher vulnerability probabilities for vulnerable versions under a common length input. During inference, the length constraint averages probabilities over fixed reference lengths. This prevents actual length from affecting the output through the explicit length input and mitigates fix reversal.
\item On PrimeVul, CFVul achieves a PairRank of 59.4\%, exceeding the existing methods compared. PairRank also improves with four additional models and three additional datasets, showing that the method mitigates fix reversal across models and datasets.
\end{itemize}
\section{Related work}\label{sec:related}

\subsection{Function-level vulnerability detection and evaluation limitations}\label{sec:rw-detection}
Deep-learning approaches typically train and evaluate vulnerability detectors at function level, using graph representation learning, pretrained code models or combinations of the two. Graph-based approaches construct code graphs from program structure or token relationships and learn function representations with graph neural networks (GNNs)~\citep{zhou2019devign,chakraborty2022deep,li2021ivdetect,nguyen2022regvd}. Pretrained approaches extract code representations using models such as CodeBERT, GraphCodeBERT, UniXcoder and CodeT5, then fine-tune them for vulnerability detection~\citep{feng2020codebert,guo2021graphcodebert,guo2022unixcoder,wang2021codet5}. Fusion approaches jointly model code sequences and program graphs~\citep{liu2025vullmgnn,bian2026taccs}. Recent studies also use large language models (LLMs) to determine directly whether functions contain vulnerabilities~\citep{steenhoek2024err,du2026vulrag}. To accommodate long functions, FuncVul performs detection using code blocks~\citep{halder2025funcvul}. Multiple-instance learning provides a basis for aggregating local code units into function-level decisions~\citep{ilse2018attention,gu2026weakly}.

Conventional classification metrics do not always reflect a detector's ability to recognise vulnerabilities. Chakraborty et al. found substantial performance declines on real-world data for models that performed well on synthetic data~\citep{chakraborty2022deep}. PrimeVul further exposed performance limitations under stricter deduplication and temporal splits~\citep{ding2025primevul}. Risse and B\"ohme found that semantics-preserving code transformations substantially reduced detection accuracy, revealing dependence on features unrelated to vulnerabilities~\citep{risse2024limits}. Risse et al. further reported that most functions in their examined data lacked the context needed for standalone vulnerability assessment~\citep{risse2025top}. These studies indicate that classification performance on individual functions cannot fully establish a detector's decision basis. Evaluation must also examine whether detectors respond consistently to vulnerability-related code changes.

\subsection{Evaluation and learning with fix pairs}\label{sec:rw-fix-pairs}
Counterfactual comparisons examine or constrain a model's decision basis by comparing outputs before and after local input changes. Contrast sets probe decision boundaries through local modifications~\citep{gardner2020contrast}. Counterfactual data augmentation provides training signals through examples whose labels change~\citep{kaushik2020learning}, while counterfactual invariance concerns stability under task-irrelevant changes~\citep{veitch2021counterfactual}. In vulnerability detection, CausalVul and VISION reduce dependence on spurious features using perturbations or counterfactual data augmentation~\citep{rahman2024causalvul,egea2025vision}. Effective fixes and irrelevant perturbations impose different output requirements. Effective fixes should lower vulnerability probability, whereas irrelevant surface changes should leave the decision unchanged. Fix pairs provide the former comparison and are already available in function-level datasets for evaluating and training detectors.

For evaluation, Risse and B\"ohme used vulnerable--patched pairs to reveal detectors' difficulty in distinguishing the two versions~\citep{risse2024limits}. Steenhoek et al. observed similar limitations in code LLMs~\citep{steenhoek2024err}. Vul-RAG introduced PairVul, a benchmark of lexically similar pairs, to evaluate discrimination between vulnerable and fixed versions~\citep{du2026vulrag}. Based on thresholded outputs, PrimeVul classified pair outcomes into four categories: both versions correct, both classified as vulnerable, both classified as non-vulnerable, and reversed classifications~\citep{ding2025primevul}. Cicho\'n and Dmitruk analysed how alarm propensity affects function-level F1. They used threshold-free pairwise ranking and length-matched analyses to examine discrimination~\citep{cichon2026flagrate}. These studies primarily assess whether detectors distinguish the versions. They give less attention to the relationship between fix-induced length changes and systematic misordering.

Fix pairs also provide supervision for training vulnerability detectors. VDSimilar learns similarity differences between vulnerable code and its patches using a Siamese bidirectional long short-term memory network (BiLSTM)~\citep{sun2021vdsimilar}. SVulD distinguishes vulnerable-function and fixed-version representations through contrastive learning with UniXcoder~\citep{ni2023svuld}. CA-SVD combines code pruning, Siamese gated graph neural networks and a contrastive loss to extract differences between the versions~\citep{song2025casvd}. RECON formulates their discrimination within a zero-sum game and incorporates prototype learning to improve detection~\citep{wen2024recon}. These methods show that fix pairs provide supervision beyond individual function labels. They primarily separate representations, without constraining the ordering of vulnerability probabilities or excluding interference from fix-induced length changes.

\subsection{Shortcut learning and control of length bias}\label{sec:rw-length}
Shortcut learning occurs when models use readily available training features that may not support generalisation to the target task~\citep{geirhos2020shortcut}. Risse et al.~\citep{risse2025top} showed that word-frequency classifiers achieved metrics close to state-of-the-art methods on common function-level datasets, indicating spurious feature--label correlations. Software size has long been studied in program analysis and defect prediction. It can confound relationships between other code metrics and defects~\citep{elemam2001confounding}, and simple size baselines can have strong discriminative ability~\citep{zhou2018progressed}. In vulnerability detection, Chakraborty et al. reported that token-based models were sensitive to input length~\citep{chakraborty2022deep}. SecVulEval observed that models detected vulnerabilities more readily in shorter functions~\citep{ahmed2025secvuleval}. However, an association between length and detection performance does not directly establish that models substitute length for vulnerability-related information. Difficulties with long inputs, dataset length--label associations and model dependence on length require separate examination.

Training-time bias control can adjust group contributions or learn relationships that remain stable across environments. Importance weighting adjusts each group's contribution to the training objective~\citep{byrd2019importance}. Group distributionally robust optimisation (Group DRO) optimises worst-group risk over predefined groups~\citep{sagawa2020distributionally}. Invariant risk minimisation (IRM) seeks representations for which the optimal classifier remains consistent across environments~\citep{arjovsky2019invariant}. These methods use distributional differences between groups, without using the correspondence between versions of the same function before and after a fix.

Another approach assigns known biases to a separate model. In natural language processing, ensemble debiasing combines the main model with a bias-only model during training. The latter absorbs the component explained by bias features; inference uses only the main model~\citep{clark2019dont,he2019unlearn}.

Length effects can also be controlled during inference. Length normalisation in information retrieval corrects the effect of document length on relevance scores~\citep{singhal2017pivoted}. Direct standardisation averages over a common reference-population distribution to make risks comparable across populations~\citep{hernan2020causal}. These approaches control bias during training or inference, without using the within-function comparison provided by fix pairs.
\section{Problem definition and motivation}\label{sec:problem}
\subsection{Fix reversal}\label{sec:phenomenon}
\paragraph{Vulnerability probability}
In function-level vulnerability detection, each sample comprises a function $x$ and a label $y\in\{0,1\}$. Here, $y=1$ denotes a vulnerable function and $y=0$ a non-vulnerable function. We denote the detector by $s:\mathcal{X}\to[0,1]$. Its output $s(x)$ is the estimated probability that a function is vulnerable, evaluated for $x$; we call it the vulnerability probability.

\paragraph{Fix pairs}
Datasets constructed from vulnerability-fixing commits retain both versions of the same function before and after a fix. These form a fix pair $(x^{+},x^{-})$, where $x^{+}$ is the vulnerable version and $x^{-}$ the fixed version. Under the assumption of fix validity, the detector should assign a higher vulnerability probability to the vulnerable version.

\paragraph{Counterfactual contrast}
A fix can be viewed as an intervention $F$ on a function, with $x^{-}=F(x^{+})$. The versions in a fix pair represent the same function before and after fixing, providing counterfactuals in the causal-inference sense~\citep{pearl2009causality}. Outcomes for the same individual under different interventions are generally not jointly observable~\citep{holland1986statistics}. Fix pairs make the two outcomes directly comparable. We call the difference between the detector's outputs the counterfactual contrast, defined as
\begin{equation}\label{eq:cf}
\Delta_s(x^{+})=s(x^{+})-s\big(F(x^{+})\big)=s(x^{+})-s(x^{-}).
\end{equation}

\paragraph{Fix reversal}
A fix pair $(x^{+},x^{-})$ exhibits fix reversal if $\Delta_s(x^{+})<0$: the fixed version receives a higher vulnerability probability than the vulnerable version. Let $\mathcal{Q}$ denote a set of fix pairs. We say that detector $s$ exhibits fix reversal on $\mathcal{Q}$ if $\mathcal{Q}$ contains any reversed pairs.

\paragraph{PairRank}
We quantify fix reversal using the fix-pair ranking metric PairRank. For a non-empty fix-pair set $\mathcal{Q}$, PairRank is the percentage of pairs in which the vulnerable version receives a higher vulnerability probability. A pair with equal vulnerability probabilities counts as half a pair:
\begin{equation}\label{eq:pairrank}
\begin{aligned}
\mathrm{PairRank}(s;\mathcal{Q})=\frac{100\%}{|\mathcal{Q}|}\sum_{(x^{+},x^{-})\in\mathcal{Q}}\Big(&\mathbb{1}\big[\Delta_s(x^{+})>0\big]\\[-2pt]
&+\tfrac12\,\mathbb{1}\big[\Delta_s(x^{+})=0\big]\Big),
\end{aligned}
\end{equation}
where $\mathbb{1}[\cdot]$ is the indicator function. Lower PairRank indicates more severe fix reversal. When PairRank is below 50\%, reversed pairs outnumber pairs in which the vulnerable version receives a higher vulnerability probability. We describe fix reversal as severe under this condition.

\subsection{Observations from the data}\label{sec:observations}
\begin{figure*}[width=\textwidth,pos=!t]
\centering
\includegraphics[width=0.92\textwidth]{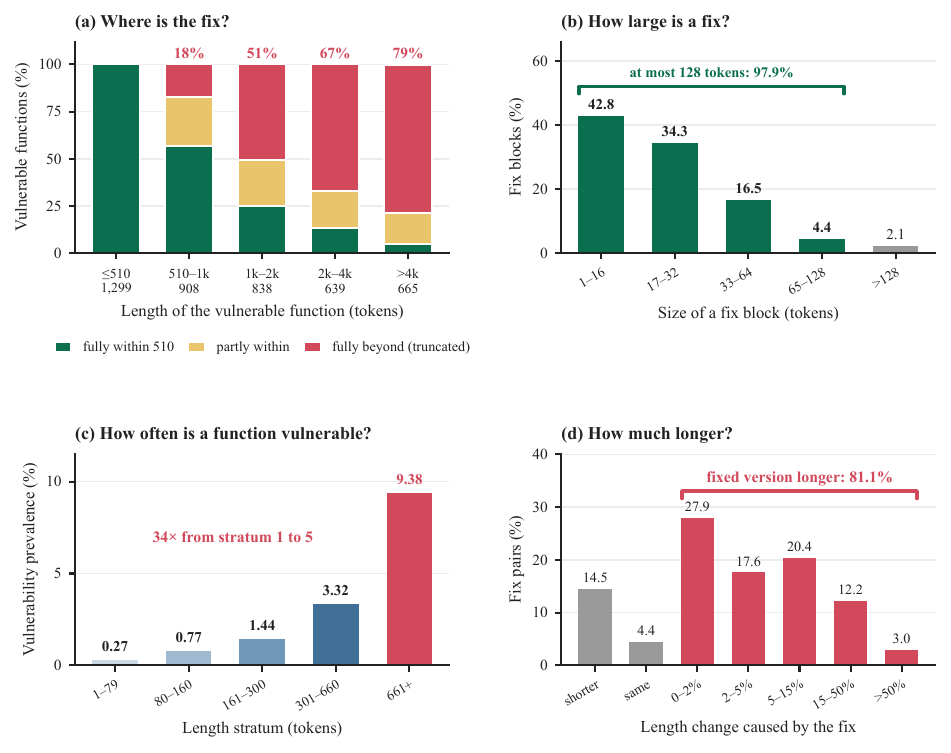}
\caption{Statistics for the PrimeVul training set. (a) Fix locations relative to the first 510 tokens, grouped by function length, for 4,349 vulnerable functions with locatable fixes. (b) Size distribution of 11,417 fix blocks. (c) Positive-class prevalence in each function-length stratum. (d) Length changes from vulnerable to fixed versions.}
\label{fig:data-coverage}
\end{figure*}

We examine the official PrimeVul~\citep{ding2025primevul} training set, with the resulting statistics shown in Figure~\ref{fig:data-coverage}.

\paragraph{Function length}
We measure function length by the number of content tokens obtained by tokenising the complete function with CodeBERT, denoted $N(x)$. Every function satisfies $N(x)\ge1$. Special tokens are excluded, and functions exceeding the model's input limit are counted at full length. We represent length using the natural logarithm of $N(x)$, written $L(x)=\log N(x)$. The difference between two log lengths equals the logarithm of their length ratio, so differences in $L$ measure relative length changes. The decomposition, explicit length input and length-related metrics below all use $L(x)$.

\paragraph{Length strata}
We divide functions into five function-length strata according to $N(x)$:
\begin{equation}\label{eq:strata}
\begin{gathered}
S_k=\{x:\tau_{k-1}<N(x)\le\tau_k\},\\
k=1,\dots,5,\quad \tau_0=0,\ \tau_5=\infty.
\end{gathered}
\end{equation}
The boundaries $\tau_1,\dots,\tau_4$ are 79, 160, 300 and 660, giving approximately equal proportions in the PrimeVul training set. The five ranges are 1--79, 80--160, 161--300, 301--660 and at least 661 tokens.

\paragraph{O1\quad Prefix truncation can omit local fixes}
With a 510-token input limit, 79\% of fixes in vulnerable training functions exceeding 4,000 tokens lie entirely beyond the prefix. These changes are invisible to prefix-only detectors, as shown in Figure~\ref{fig:data-coverage}a. Fixes themselves are usually small: 97.9\% of fix blocks contain at most 128 tokens, as shown in Figure~\ref{fig:data-coverage}b.

\paragraph{O2\quad Training-set positive-class prevalence increases with length}
Positive-class prevalence rises from 0.27\% in the shortest training-set stratum to 9.38\% in the longest, as shown in Figure~\ref{fig:data-coverage}c. Length therefore discriminates between vulnerable and non-vulnerable functions in the training set.

\paragraph{O3\quad Fixes usually make functions longer}
In 81.1\% of training fix pairs, the fixed version is longer than the vulnerable version, as shown in Figure~\ref{fig:data-coverage}d. A rule that assigns vulnerability probabilities solely by increasing length would reverse all these pairs.

\subsection{Length shortcut and task definition}\label{sec:essence}
\paragraph{Content and length terms}
A fix changes both vulnerability-related code and function length. The counterfactual contrast $\Delta_s$ therefore combines their effects, which we separate below. We write the detector as $s(x)=s\big(c(x),L(x)\big)$, where $c(x)$ denotes code content. For every actual function, $s(c(x),L(x))=s(x)$. The vulnerable version has content and length $(c^{+},L^{+})$, and the fixed version has $(c^{-},L^{-})$. The quantity $s(c^{+},L^{-})$ is the vulnerability probability for the vulnerable content evaluated at the fixed version's length. Adding and subtracting $s(c^{+},L^{-})$ in Equation~\eqref{eq:cf} gives
\begin{equation}\label{eq:decomp}
\begin{aligned}
\Delta_s(x^{+})={}&\underbrace{s(c^{+},L^{-})-s(c^{-},L^{-})}_{\Delta_s^{\mathrm{cont}}}\\
&+\underbrace{s(c^{+},L^{+})-s(c^{+},L^{-})}_{\Delta_s^{\mathrm{len}}}.
\end{aligned}
\end{equation}
The content term $\Delta_s^{\mathrm{cont}}$ fixes length at $L^{-}$ and changes content from $c^{+}$ to $c^{-}$. The length term $\Delta_s^{\mathrm{len}}$ fixes content at $c^{+}$ and changes length from $L^{+}$ to $L^{-}$.

\paragraph{Length shortcut}
For given code content $c$, a detector exhibits a length shortcut if $s(c,L)$ increases strictly with $L$. The term derives from shortcut learning~\citep{geirhos2020shortcut}. When a detector has an explicit length input, setting that input directly gives $s(c,L)$. Otherwise, we increase length by inserting semantics-preserving comments. The resulting change in vulnerability probability provides an operational definition of the length shortcut.

\paragraph{Conditions for fix reversal in an individual pair}
An individual fix pair exhibits fix reversal when all three conditions below hold.
\begin{enumerate}
\item A length shortcut exists: vulnerability probability increases strictly with length when content is fixed, so $s(c^{+},L)$ increases strictly with $L$.
\item The fix makes the function longer, so $L^{-}>L^{+}$. Together with Condition 1, this implies a negative length term, $\Delta_s^{\mathrm{len}}<0$.
\item The length term outweighs the content term: $|\Delta_s^{\mathrm{len}}|>\Delta_s^{\mathrm{cont}}$.
\end{enumerate}
Equation~\eqref{eq:decomp} then gives $\Delta_s(x^{+})=\Delta_s^{\mathrm{cont}}+\Delta_s^{\mathrm{len}}<0$. Fix reversal becomes more severe as more pairs in the set satisfy these three conditions.

\paragraph{Problem and task requirements}
We aim to weaken the length shortcut and thereby mitigate fix reversal. Training may use labelled functions and fix pairs, while inference receives only a single function. This task imposes two requirements on the detector. Following O1, Requirement 1 is complete-function coverage. If a fix lies beyond the truncation boundary, the visible content is identical, the content term is 0, and the fix cannot inform the ordering. Following O2 and O3, Requirement 2 is control of length effects. During training, within-pair length differences should not cause the fixed version to receive a higher vulnerability probability. During inference, a function's vulnerability probability should not change with its actual length.

\section{Method}\label{sec:method}
CFVul comprises sliding-window encoding A, content constraint B and length constraint C. Component A addresses Requirement 1 in Section~\ref{sec:essence}. Components B and C address the fix-pair and individual-function aspects of Requirement 2, respectively.

\subsection{Sliding-window encoding}\label{sec:compA}\label{sec:head}
Sliding-window encoding covers complete functions, exposing fix changes in long functions. It comprises window segmentation and attention pooling. The function body is divided into $m$ windows, each containing at most $W$ tokens. Window boundaries align with code-line boundaries; a line is split internally only when it exceeds $W$ tokens. Adjacent window starts are separated by at most $W/2$ tokens, and the windows collectively cover the complete body. We set $W$ to 128, covering the size of 97.9\% of fix blocks in O1. Each window is prefixed with the function signature, comprising its return type, name and parameter list, to retain function-level context. A shared encoder maps window $j$ to representation $e_j$. Attention pooling computes window weights using tanh attention~\citep{ilse2018attention} and obtains the function representation as a weighted sum. The weights and function representation are
\begin{equation}\label{eq:attn}
\begin{gathered}
u_j=w_{\mathrm{att}}^\top\tanh\!\big(V_{\mathrm{att}}e_j+b_{\mathrm{att}}\big),\\
\alpha_j=\frac{\exp(u_j)}{\sum_{t=1}^{m}\exp(u_t)},\qquad z(x)=\sum_{j=1}^{m}\alpha_j e_j.
\end{gathered}
\end{equation}
Here, $e_j\in\mathbb{R}^{d}$ and the window-representation dimension $d$ is 768. The learnable parameters $V_{\mathrm{att}}\in\mathbb{R}^{h\times d}$, $b_{\mathrm{att}}\in\mathbb{R}^{h}$ and $w_{\mathrm{att}}\in\mathbb{R}^{h}$ are shared across all functions and windows. The attention mechanism has hidden dimension $h$ and assigns score $u_j$ to window $j$. Normalising its weight $\alpha_j$ across the $m$ windows of the same function gives $\sum_{j=1}^{m}\alpha_j=1$, yielding the function representation $z(x)\in\mathbb{R}^{d}$.

The classification head receives the function representation $z(x)$ and function length. It is defined as $f(z,L)=\mathrm{MLP}([z;\phi(L)])$, where $\mathrm{MLP}$ is a multilayer perceptron (MLP) and $\phi$ is a log-length embedding. The separate length input $L$ can be set independently of the code representation; we call it the explicit length input. During training, this input takes the function's actual length, giving logit $f(z(x_i),L(x_i))$ for function $x_i$. The classification head is trained with the function-level binary cross-entropy (BCE) loss:
\begin{equation}\label{eq:bcew}
\begin{aligned}
\Lbce=-\frac{1}{n}\sum_{i=1}^{n}\Big[&y_i\log\sigma\!\big(f(z(x_i),L(x_i))\big)\\
&+(1-y_i)\log\big(1-\sigma\!\big(f(z(x_i),L(x_i))\big)\big)\Big].
\end{aligned}
\end{equation}
The classification head retains an explicit length input because vulnerability prevalence increases with length in the training set. Using the actual length gives vulnerability probability $\sraw(x)=\sigma\big(f(z(x),L(x))\big)$, termed the actual-length output. The explicit length input also provides a separately controllable entry point for length. Both the content and length constraints operate by setting its value. The function representation $z(x)$ corresponds to code content $c(x)$ in Equation~\eqref{eq:decomp}.

\subsection{Content constraint}\label{sec:pairrank}\label{sec:compB}\label{sec:train-deploy}
The content constraint addresses the fix-pair aspect of Requirement 2 and the content term in Equation~\eqref{eq:decomp}. During training, within-pair length differences should not cause higher vulnerability probability for the fixed version. For each training fix pair $(x^{+},x^{-})$, the constraint fixes both explicit length inputs at their mean log length $\bar L$ and encourages a higher vulnerability probability for the vulnerable version. The pairwise logit difference is
\begin{equation}\label{eq:pair-contrast}
\begin{gathered}
\Delta_f(x^{+};\bar L)=f\big(z(x^{+}),\bar L\big)-f\big(z(x^{-}),\bar L\big),\\
\bar L=\frac{L(x^{+})+L(x^{-})}{2}.
\end{gathered}
\end{equation}
The content constraint uses the Bradley--Terry paired-comparison model~\citep{bradley1952rank}, with $\exp(f(z(x),\bar L))$ as a positive-valued strength. The pairwise ranking probability that the vulnerable version ranks above the fixed version is
\begin{equation}\label{eq:bt-pair}
\begin{aligned}
p_{\mathrm{BT}}(x^{+},x^{-};\bar L)
&=\frac{\exp\!\big(f(z(x^{+}),\bar L)\big)}{\exp\!\big(f(z(x^{+}),\bar L)\big)+\exp\!\big(f(z(x^{-}),\bar L)\big)}\\
&=\sigma\!\big(\Delta_f(x^{+};\bar L)\big).
\end{aligned}
\end{equation}
This pairwise ranking probability describes the relative ordering of the two versions, while an individual function's vulnerability probability is given by the classification head's sigmoid output. The training functions used in one parameter update constitute an optimisation group $G$. Let $\mathcal{P}_G$ denote the fix pairs whose two versions both occur in $G$. The content constraint minimises the mean negative log-likelihood (NLL) of the ranking probability over $\mathcal{P}_G$, termed the pairwise loss:
\begin{equation}\label{eq:pair}
\begin{aligned}
\mathcal{L}_{\mathrm{pair}}
&=-\frac{1}{|\mathcal{P}_G|}\sum_{(x^{+},x^{-})\in\mathcal{P}_G}\log p_{\mathrm{BT}}(x^{+},x^{-};\bar L)\\
&=\frac{1}{|\mathcal{P}_G|}\sum_{(x^{+},x^{-})\in\mathcal{P}_G}\log\!\Big(1+\exp\!\big(-\Delta_f(x^{+};\bar L)\big)\Big).
\end{aligned}
\end{equation}
If a group contains no complete pair, this loss term is $0$. The expression is a logistic pairwise loss, a smooth surrogate for pairwise ranking errors. It encourages $\Delta_f>0$, giving the vulnerable version a higher vulnerability probability. Both versions share explicit length input $\bar L$, so their actual length difference does not enter the comparison through this input.

Function-level datasets retaining both versions and their correspondence provide the required supervision directly from existing fix pairs, without additional commit collection or annotation. The versions belong to the same project and function, share most code, and differ mainly in fix-related changes. Comparing their vulnerability probabilities therefore avoids differences in project and function background. Because most training fix pairs have longer fixed versions, a detector that assigns higher vulnerability probabilities on the basis of length incurs a larger pairwise loss.

The overall training objective of CFVul is
\begin{equation}\label{eq:total}
\begin{aligned}
\mathcal{L}=\Lbce&+\lambda_{\mathrm{inv}}\,\mathrm{ERM}_{\mathrm{inv}}\\
&+\lambda_{\mathrm{aux}}\,\Laux+\lambda_{\mathrm{pair}}\,\mathcal{L}_{\mathrm{pair}},
\end{aligned}
\end{equation}
where $\mathrm{ERM}_{\mathrm{inv}}$ and $\Laux$ are auxiliary regularisers used only during training and excluded from inference. The term $\mathrm{ERM}_{\mathrm{inv}}$ is the empirical risk minimisation (ERM) loss of an auxiliary invariant head $v^\top z+b_e$. It is computed separately in four length environments formed by merging function-length strata 1 and 2 and retaining the other three strata. All environments share a common slope $v$ and use separate intercepts $b_e$, allowing only the intercept to vary across environments. This design draws on invariant risk minimisation (IRM)~\citep{arjovsky2019invariant}. The term $\Laux$ is the loss of a linear window classifier that distinguishes windows changed by the fix in the vulnerable version from corresponding windows in the fixed version. Appendix Table~\ref{tab:hparams} gives the weights and other hyperparameters. The pairwise loss requires both versions in the same optimisation group. Gradient caching~\citep{gao2021scaling} handles the memory cost of long functions.

\subsection{Length constraint}\label{sec:compC}
\begin{figure*}[width=\textwidth,pos=!t]
\centering
\includegraphics[width=\textwidth]{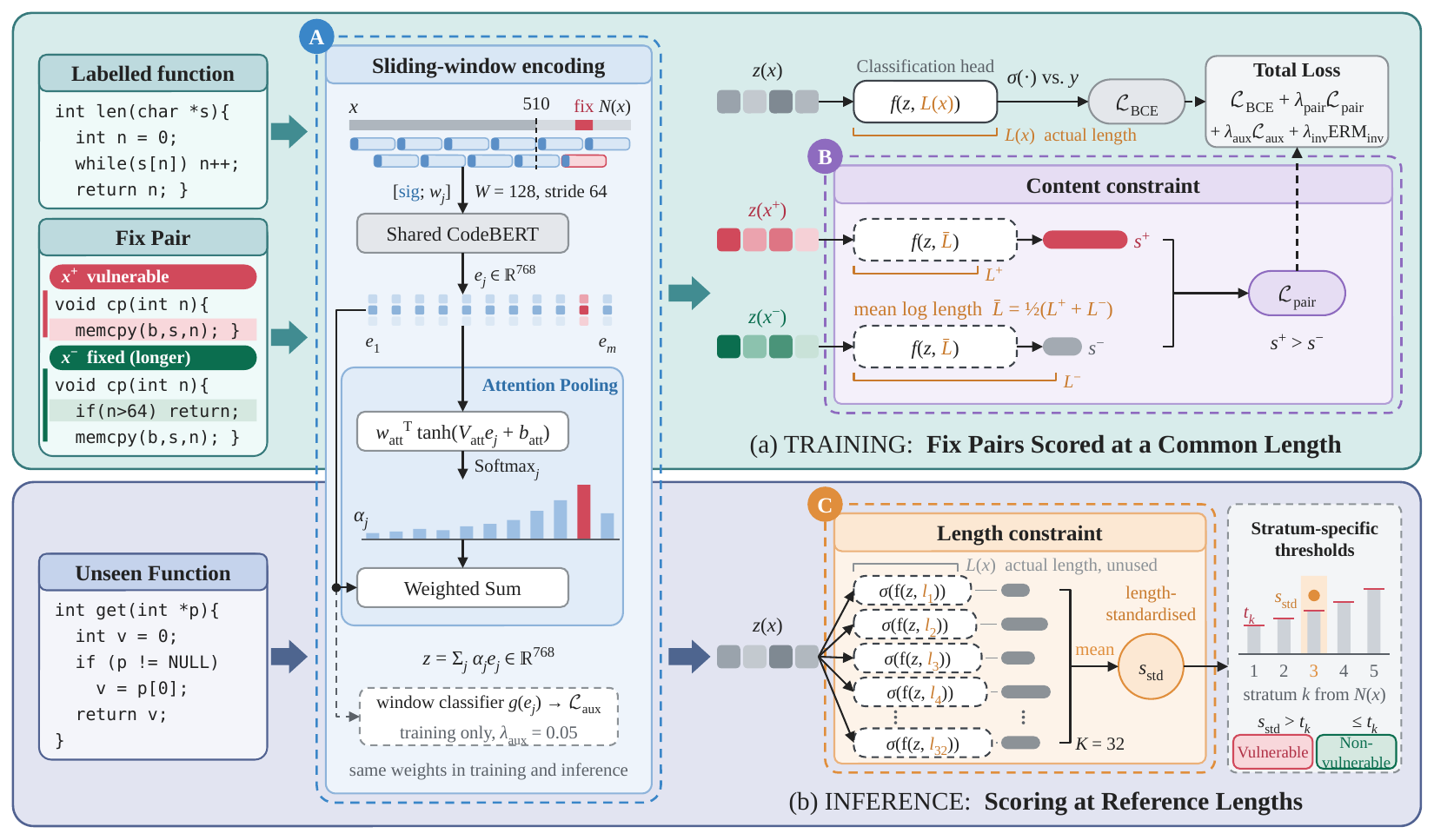}
\caption{Training and inference in CFVul. (a) During training, BCE uses the actual length for every function, while the pairwise loss uses a common length $\bar L$ for fix pairs. (b) During inference, $\Kref$ reference lengths yield $\sstd$, and stratum-specific thresholds determine alarms.}
\label{fig:model}
\end{figure*}

The length constraint addresses the individual-function aspect of Requirement 2: inference vulnerability probability should not change with actual function length. CFVul takes only one function at inference, and actual length raises long functions' vulnerability probabilities through the explicit length input. The length constraint substitutes fixed reference lengths for actual length, blocking this route of dependence. We draw $K$ reference log lengths $l_1,\dots,l_K$ once from the training-set length distribution. The resulting vulnerability probability, termed the length-standardised output, is
\begin{equation}\label{eq:sstd}
s_{\mathrm{std}}(x)=\frac{1}{K}\sum_{r=1}^{K}\sigma\bigl(f(z(x),l_r)\bigr).
\end{equation}
With a fixed reference set, $s_{\mathrm{std}}$ is a function only of $z(x)$, and every function uses the same reference-length set. The code and function representation remain fixed while the explicit length input is set to each reference length, and the resulting probabilities are averaged. This follows the computational form of backdoor adjustment in causal inference: averaging outputs over a reference-length distribution~\citep{pearl2009causality}. CausalVul~\citep{rahman2024causalvul} similarly averages over spurious features such as variable names in vulnerability detection. We refer to the content and length constraints collectively as counterfactual constraints. At deployment, CFVul uses $s_{\mathrm{std}}$ with stratum-specific thresholds to control the false positive rate (FPR) in each stratum. Section~\ref{sec:rq3} describes threshold selection. Figure~\ref{fig:model} illustrates the training and inference procedures.

\subsection{Evaluation metrics}\label{sec:metrics}
Conventional classification metrics comprise overall AUROC and precision, recall and F1 for the vulnerable class. The classification threshold maximises validation F1. Only 695 of the 25,911 PrimeVul test functions are vulnerable, indicating strong class imbalance; see Table~\ref{tab:datasets}. We therefore exclude accuracy from the conventional classification metrics. With a length--label association, a detector using a length shortcut can achieve high overall AUROC. This metric cannot distinguish reliance on code content from reliance on length. Conventional classification metrics thus serve only for comparison with prior work.

We use two primary metrics at each of two levels: fix pairs and individual functions. At fix-pair level, PairRank and $\rho_{\Delta}$ are computed on test fix pairs. At individual-function level, functions are evaluated separately to compute length-matched AUROC and $\rho_{L}$. All metrics use the detector's vulnerability probability $s(x)$; CFVul uses the length-standardised output $\sstd(x)$. The 95\% bootstrap confidence intervals for fix-pair metrics are obtained by paired bootstrap resampling of test fix pairs.

PairRank is computed using Equation~\eqref{eq:pairrank}, with $\mathcal{Q}$ denoting test fix pairs unless otherwise stated. Lower PairRank indicates more severe fix reversal.
The metric $\rho_{\Delta}$ examines the fix-pair aspect of Requirement 2, corresponding to the content constraint. Within-pair length differences should not cause higher vulnerability probability for the fixed version. We define it as
\begin{equation}\label{eq:rhopair}
\begin{aligned}
\rho_{\Delta}(s)=\mathrm{Spearman}_{(x^{+},x^{-})\in\mathcal{Q}}\Big(&L(x^{-})-L(x^{+}),\\[-2pt]
&s(x^{-})-s(x^{+})\Big),
\end{aligned}
\end{equation}
where $L(x^{-})-L(x^{+})=\log\big(N(x^{-})/N(x^{+})\big)$ is the natural logarithm of the fixed-to-vulnerable length ratio. Ties receive average ranks. The range of $\rho_{\Delta}$ is $-1$ to $1$. A positive value indicates that larger fix-induced length increases accompany relatively higher vulnerability probabilities for fixed versions. Values closer to 0 indicate less influence of length changes on pairwise ordering.

Length-matched AUROC measures discrimination between vulnerable and non-vulnerable functions of similar length based on code content. For the $k$th function-length stratum $S_k$, we define the set of matched vulnerable/non-vulnerable pairs and the stratum's length-matched AUROC as
\begin{equation}\label{eq:samelen}
\begin{aligned}
P_k=\Big\{(i,j):\ &y_i=1,\ x_i\in S_k,\\[-2pt]
&y_j=0,\ \big|L(x_i)-L(x_j)\big|\le0.05\Big\},\\
R_k(s)=\frac{1}{|P_k|}\sum_{(i,j)\in P_k}\Big[&\mathbb{1}\big(s(x_i)>s(x_j)\big)\\[-2pt]
&+\tfrac12\,\mathbb{1}\big(s(x_i)=s(x_j)\big)\Big].
\end{aligned}
\end{equation}
A log-length difference of at most 0.05 corresponds to a length ratio approximately within $[0.951,1.051]$. Matched non-vulnerable functions may come from adjacent strata; vulnerable functions without a match are excluded. Length-matched AUROC is the arithmetic mean across five strata, $\overline R(s)=\tfrac15\sum_{k=1}^{5}R_k(s)$, reported only when all five have matched pairs. Its range is 0 to 1. A value above 0.5 indicates that the vulnerable function more often receives the higher vulnerability probability among matched pairs. Higher values indicate stronger discrimination based on code content. This metric checks whether detection ability is retained when fix reversal is mitigated.

The metric $\rho_{L}$ examines the individual-function aspect of Requirement 2, corresponding to the length constraint. A function's vulnerability probability should not change with its actual length. Let $\mathcal{T}_0$ denote the non-vulnerable test functions. We define
\begin{equation}\label{eq:rholen}
\rho_{L}(s)=\mathrm{Spearman}_{x\in\mathcal{T}_0}\Big(L(x),\ s(x)\Big).
\end{equation}
Vulnerable functions are generally longer, so this calculation uses only non-vulnerable functions to exclude the label--length association from length dependence. The range of $\rho_{L}$ is $-1$ to $1$. Positive values indicate higher vulnerability probabilities for longer non-vulnerable functions, while values closer to 0 indicate weaker length dependence.

\section{Experimental results and analysis}\label{sec:results}
Except in cross-model experiments, CFVul is applied to CodeBERT throughout. In figures and tables, CFVul denotes the detector obtained by applying the method.
We use four public function-level vulnerability datasets: PrimeVul~\citep{ding2025primevul}, BigVul~\citep{fan2020bigvul}, ReVeal~\citep{chakraborty2022deep} and MegaVul~\citep{ni2024megavul}. PrimeVul is the primary dataset. Its official split facilitates comparison with prior work. Stricter relabelling of vulnerable functions and removal of duplicates reduce interference from label noise and repeated samples in fix-reversal analysis. We compute the ratio of vulnerability prevalence in the longest to the shortest function-length stratum of the training set. PrimeVul has the largest ratio among the four datasets, at 34.2; see Table~\ref{tab:datasets}. The other datasets differ in this ratio and in the proportion of fixed versions included as training negatives. They test whether the conclusions hold under different data conditions.
All four datasets are constructed from vulnerability-fixing commits and contain fix pairs and unpaired functions, both with vulnerability labels. Training fix pairs contribute to classification training and supervise the content constraint. Validation and test pairs are used to compute PairRank and $\rho_{\Delta}$, assessing whether vulnerable versions receive higher vulnerability probabilities. Unpaired functions have no corresponding fixed version and constitute most functions in each dataset. They join both versions of fix pairs in classifier training and contribute to test metrics such as length-matched AUROC.
PrimeVul uses its official split. The remaining datasets are split by fixing commit into training, validation and test sets at $8:1:1$, keeping both versions of a pair together. Table~\ref{tab:datasets} reports dataset sizes; Appendix~\ref{app:repro} describes deduplication and cleaning.

\begin{table*}[width=\textwidth,pos=!t]
\centering
\caption{Datasets used in the experiments. Split ratio gives the planned training / validation / test proportions. PrimeVul uses its official split; the other datasets are grouped by fixing commit. Train, Valid and Test give total / vulnerable function counts. Fix pairs gives training / validation / test pair counts. In Dedup., official denotes the dataset's official deduplication. Fixed versions records the source of fixed-version negatives: original means already present, and added means recovered from fixing commits in this study. Fixed included is the proportion of vulnerable training functions whose fixed versions are also included as negative examples. Vul.-rate ratio is the ratio of vulnerability prevalence in the longest to the shortest function-length stratum of the training set.}
\label{tab:datasets}
\small
\setlength{\tabcolsep}{3.6pt}
\begin{tabular}{@{}lccccccccc@{}}
\hline
Dataset & \shortstack{Split\\ratio} & Train & Valid & Test & \shortstack{Fix pairs\\train / valid / test} & Dedup. & \shortstack{Fixed\\versions} & \shortstack{Fixed\\included} & \shortstack{Vul.-rate\\ratio} \\
\hline
PrimeVul & 8:1:1 & 184,427 / 5,574 & 25,430 / 699 & 25,911 / 695 & 4,354 / 562 / 564 & official & original & 78\% & 34.2$\times$ \\
BigVul & 8:1:1 & 155,575 / 8,407 & 21,242 / 1,227 & 22,469 / 1,266 & 8,229 / 1,199 / 1,222 & no & added & 98\% & 6.1$\times$ \\
ReVeal & 8:1:1 & 16,771 / 1,452 & 2,769 / 237 & 2,492 / 200 & 1,369 / 197 / 195 & yes & original & 94\% & 7.0$\times$ \\
MegaVul & 8:1:1 & 284,762 / 14,076 & 34,305 / 1,646 & 36,537 / 1,853 & 12,768 / 1,500 / 1,712 & yes & added & 91\% & 13.4$\times$ \\
\hline
\end{tabular}
\end{table*}

\subsection{RQ1: Fix reversal in existing detectors and its causes}

\subsubsection{Fix reversal and its relationship with length}\label{sec:rq1-reversal}\label{sec:rq1-free}
Existing detectors widely exhibit severe fix reversal. On PrimeVul's 564 test fix pairs, CodeBERT, UniXcoder, GraphCodeBERT and CodeT5 all had PairRank below 50\%. Of these 564 pairs, 475 had longer fixed versions. This agrees with the training-set observation that most fixes lengthen functions, reported in Section~\ref{sec:observations}.

\begin{figure}[pos=!t]
\centering
\includegraphics[width=\linewidth]{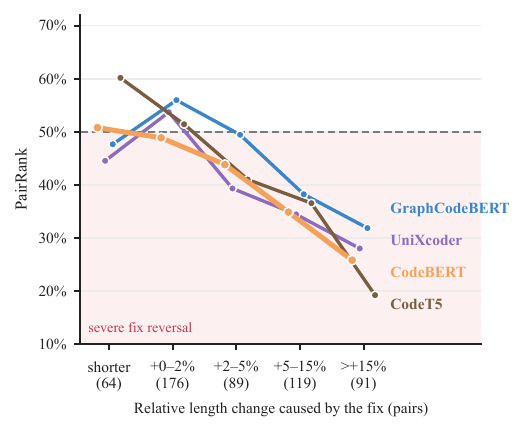}
\caption{PairRank of four detectors on test fix pairs grouped by the relative length change caused by the fix. The 25 equal-length pairs are excluded. Table~\ref{tab:rq1-dose} reports group values and 95\% bootstrap confidence intervals.}
\label{fig:rq1-dose}
\end{figure}

Larger fix-induced length increases were associated with more severe fix reversal. We grouped test pairs by the relative length change caused by the fix. Among groups with longer fixed versions, all four detectors' PairRank decreased as the length increase grew; see Figure~\ref{fig:rq1-dose}. Fix reversal thus concentrated in pairs with substantially longer fixed versions. In these pairs, length and content at the fix location change together, so grouping by length cannot separate their effects. Section~\ref{sec:rq1-cause} therefore uses code transformations to control length and content separately and examine whether length dependence explains fix reversal.

\subsubsection{Code-transformation experiments}\label{sec:rq1-cause}
We transformed the vulnerable versions in fix pairs and recomputed their vulnerability probabilities using the four trained detectors. Two experiment groups examined the role of length from complementary directions. Length padding preserves program content while increasing length, testing whether PairRank changes. The equal-length comparison matches version lengths to test whether fix content alone supports higher vulnerability probabilities for vulnerable versions. We selected PrimeVul validation and test pairs with longer fixed versions and all changes within the first 500 tokens. Of these, 133 pairs admitted exact length matching under all four tokenisers, with every transformed version within 510 tokens. They comprised 74 validation and 59 test pairs. Before transformation, PairRank was 12\%, 28\%, 22\% and 19\% for CodeBERT, UniXcoder, GraphCodeBERT and CodeT5, respectively. Reversal was more severe than on all test pairs; see Appendix Figure~\ref{fig:iv-fillers}.

\begin{figure*}[width=\textwidth,pos=!t]
\centering
\includegraphics[width=\textwidth]{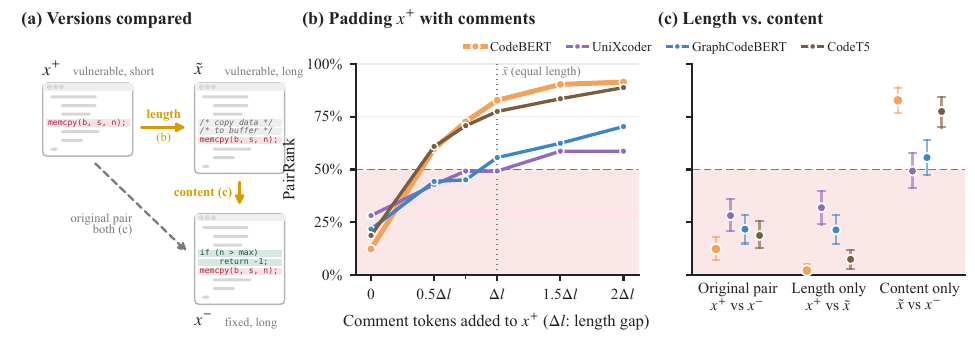}
\caption{Code-transformation experiments. (a) The three versions compared. (b) Length padding. (c) Equal-length comparison. Error bars show 95\% bootstrap confidence intervals.}
\label{fig:iv-main}
\end{figure*}

Both experiments transformed only the vulnerable version, inserting semantics-preserving padding at the fix location. Denote the vulnerable version by $x^{+}$, the fixed version by $x^{-}$, and their length difference by $\Delta l$. The main-text filler is sampled real comments: sentences randomly sampled from existing comments in non-vulnerable PrimeVul training functions. Sentences mentioning checks, errors, bounds, null pointers, safety or fixes were removed beforehand. The sampled sentences were unrelated to the transformed function and vulnerability. We also tested three alternative fillers: comments generated by Qwen3-4B-Instruct-2507, unused variable declarations and hyphen-only comments. Appendix~\ref{app:rq1} reports these results. Adding $\Delta l$ comment tokens to $x^{+}$ yields a version matching $x^{-}$ in length under each detector's tokeniser, denoted $\tilde{x}$. The difference between $\tilde{x}$ and $x^{+}$ is the inserted comments. The length of $\tilde{x}$ equals that of $x^{-}$, allowing a comparison of fix content. Figure~\ref{fig:iv-main}(a) shows the three versions. Every comparison uses PairRank. When comparing $x^{+}$ with $\tilde{x}$, the latter replaces $x^{-}$. Intervals use 5,000 paired bootstrap resamples.

\paragraph{Length padding}
Length padding tests whether increasing length while preserving program content changes the ordering of versions, providing an operational test for the length shortcut. We inserted sampled real comments at the fix location in $x^{+}$, using 0, 0.5, 0.75, 1, 1.5 and 2 times $\Delta l$. Adding $\Delta l$ makes the versions equally long. This is the comparison between $\tilde{x}$ and $x^{-}$ in the equal-length experiment. Adding $2\Delta l$ makes $x^{+}$ longer than $x^{-}$ by $\Delta l$, swapping their length ordering. After padding $x^{+}$ by $2\Delta l$, the four detectors' PairRank increased from 12\%--28\% to 59\%--91\%. Values were 91\%, 59\%, 70\% and 89\% for CodeBERT, UniXcoder, GraphCodeBERT and CodeT5, respectively; see Figures~\ref{fig:iv-main}(b) and~\ref{fig:iv-fillers}. In the control with identical padding added to both versions, which preserved the length difference, all four detectors remained in the severe fix-reversal region. The rise in PairRank therefore came from the changed length difference; see Appendix~\ref{app:rq1}. All four detectors exhibited a length shortcut with sampled real comments. UniXcoder showed the weakest length effect, with the lowest PairRank, 59\%, after swapping the length ordering.

\paragraph{Equal-length comparison}
The equal-length experiment decomposes the original comparison of $x^{+}$ and $x^{-}$ into two steps to separate length and content contributions. The comparison of $x^{+}$ with $\tilde{x}$ isolates padding-induced length changes. The comparison of $\tilde{x}$ with $x^{-}$ matches length and compares fix content. The original, length-only and content-only comparisons form the three columns in Figure~\ref{fig:iv-main}(c). With sampled real comments, all four detectors assigned higher vulnerability probability to the padded $\tilde{x}$ in most length-only pairs. This occurred in almost every pair for CodeBERT and CodeT5. In the content-only comparison, their PairRank increased from 12\% and 19\% to 83\% and 77\%, respectively. GraphCodeBERT and UniXcoder increased from 22\% and 28\% to 56\% and 49\%; see Figures~\ref{fig:iv-main}(c) and~\ref{fig:iv-fillers}. Under this padding condition, CodeBERT and CodeT5 ranked vulnerable versions higher in most equal-length pairs. GraphCodeBERT and UniXcoder approached 50\% PairRank. These results support length differences as the main source of fix reversal in the examined pairs. Appendix~\ref{app:rq1} reports results for other fillers.

RQ1 shows widespread severe fix reversal among existing detectors. All four detectors had test-pair PairRank below 50\%, with more severe reversal for larger fix-induced length increases. Padding only the vulnerable version with sampled real comments until it exceeded the fixed version increased PairRank from 12\%--28\% to 59\%--91\%. Padding to equal length increased PairRank to 49\%--83\%; see Figure~\ref{fig:iv-main} and Appendix Figure~\ref{fig:iv-fillers}. Under this filler condition, the code-transformation results support the length shortcut as the main source of fix reversal in the examined pairs.

\subsection{RQ2: Effects of CFVul on the length shortcut and fix reversal}

\subsubsection{Comparison with trivial baselines and existing methods}\label{sec:baselines}
\begin{table*}[width=\textwidth,pos=!t]
\centering
\figtablefonts\footnotesize\setlength{\tabcolsep}{4pt}\renewcommand{\arraystretch}{0.88}
\caption{Comparison of CFVul with trivial baselines and existing methods on the PrimeVul test set. CFVul is shown in bold.}
\label{tab:main-algorithms}
\begin{tabular}{@{}lcccccccc@{}}
\toprule
 & \multicolumn{4}{c}{Length-aware metrics (primary)} & \multicolumn{4}{c}{Conventional metrics} \\
\cmidrule(lr){2-5}\cmidrule(l){6-9}
Method & PairRank (\%)$\uparrow$ & \shortstack{Length-matched\\AUROC$\uparrow$} & $\rho_{\Delta}$ & $\rho_{L}$ & AUROC & Precision & Recall & F1 \\
\midrule
\multicolumn{9}{@{}l}{\emph{Trivial baselines}} \\
Length only (no code) & 13.6 & 0.502 & $+1.00$ & 1.00 & 0.813 & 0.153 & 0.317 & 0.206 \\
Bag of words (TF-IDF + LR) & 36.3 & 0.647 & $+0.34$ & 0.79 & 0.848 & 0.214 & 0.353 & 0.266 \\
\midrule
\multicolumn{9}{@{}l}{\emph{Pre-trained sequence models}} \\
CodeBERT & 41.9 & 0.670 & $+0.32$ & 0.85 & 0.836 & 0.181 & 0.296 & 0.225 \\
UniXcoder & 42.7 & 0.687 & $+0.28$ & 0.67 & 0.826 & 0.177 & 0.327 & 0.229 \\
GraphCodeBERT & 45.6 & 0.710 & $+0.22$ & 0.76 & 0.834 & 0.183 & 0.288 & 0.223 \\
CodeT5 & 42.3 & 0.683 & $+0.36$ & 0.79 & 0.837 & 0.192 & 0.276 & 0.226 \\
\midrule
\multicolumn{9}{@{}l}{\emph{NCS--CPG fusion (CodeBERT + GNN)}$^\dagger$} \\
Vul-LMGNN (CodeBERT) & 36.6 & 0.643 & $+0.38$ & 0.91 & 0.843 & 0.194 & 0.189 & 0.192 \\
TaCCS-DFA & 37.2 & 0.568 & $+0.31$ & 0.84 & 0.826 & 0.131 & 0.330 & 0.188 \\
\midrule
\multicolumn{9}{@{}l}{\emph{Contrastive learning on fix pairs}} \\
SVulD (R-Drop) & 37.7 & 0.705 & $+0.42$ & 0.78 & 0.850 & 0.200 & 0.351 & 0.255 \\
SVulD (SimCL) & 39.7 & 0.702 & $+0.38$ & 0.81 & 0.851 & 0.200 & 0.364 & 0.258 \\
\midrule
\multicolumn{9}{@{}l}{\emph{Length-robust training (CodeBERT backbone)}} \\
Stratum-private & 42.6 & 0.615 & $+0.22$ & 0.90 & 0.823 & 0.184 & 0.286 & 0.224 \\
Stratum-weighted & 39.2 & 0.667 & $+0.36$ & 0.86 & 0.834 & 0.178 & 0.291 & 0.221 \\
Group DRO & 44.1 & 0.623 & $+0.17$ & 0.58 & 0.797 & 0.186 & 0.263 & 0.218 \\
\midrule
\multicolumn{9}{@{}l}{\emph{This work (two scores of one checkpoint)}} \\
\textbf{CFVul, length-standardised output ($\sstd$)} & \textbf{59.4} & \textbf{0.675} & $\mathbf{-0.05}$ & \textbf{0.67} & \textbf{0.806} & \textbf{0.180} & \textbf{0.209} & \textbf{0.193} \\
CFVul, actual-length output ($\sraw$) & 39.4 & 0.674 & $+0.40$ & 0.88 & 0.836 & 0.202 & 0.265 & 0.229 \\
\bottomrule
\end{tabular}
\\
{\footnotesize $^\dagger$ Evaluation uses only the 433 test pairs with CPGs for both versions and 24,788 test functions with CPGs. CFVul's PairRank on these 433 pairs is 60.6\%.}
\end{table*}
We compared CFVul with two trivial baselines and four categories of existing methods. Following Cicho\'n and Dmitruk~\citep{cichon2026flagrate}, the trivial baselines comprise a length-only rule and a bag-of-words baseline (BoW). The latter trains logistic regression (LR) on term frequency--inverse document frequency (TF-IDF) features of identifiers and symbols. The four method categories represent established approaches. Pretrained sequence models include CodeBERT~\citep{feng2020codebert}, UniXcoder~\citep{guo2022unixcoder}, GraphCodeBERT~\citep{guo2021graphcodebert} and CodeT5~\citep{wang2021codet5}, representing standard function-level fine-tuning. CodeBERT reads the first 510 content tokens and is the encoder used when applying CFVul in this comparison. Structural fusion methods comprise TaCCS-DFA~\citep{bian2026taccs} and Vul-LMGNN~\citep{liu2025vullmgnn}, both reproduced with CodeBERT backbones. They combine natural code sequences (NCS) and code property graphs (CPG), incorporating program structure. Fix-pair contrastive learning is represented by SVulD~\citep{ni2023svuld} in its R-Drop and SimCL variants. Both retain the original UniXcoder encoder and, like CFVul, use fix pairs during training. Length-robust training methods use CodeBERT and comprise stratum-weighted training, stratum-private training and Group DRO~\citep{sagawa2020distributionally}. These respectively reweight losses by function-length stratum, train a separate model per stratum, and optimise the worst-stratum loss. Like CFVul, all three account for function length during training.

All neural detectors followed a common training protocol, detailed in Appendix~\ref{app:repro}. TaCCS-DFA and Vul-LMGNN require code property graphs generated by Joern~\citep{yamaguchi2014cpg}. They were evaluated on 24,788 test functions with successfully generated CPGs and 433 test pairs with CPGs for both versions.

CFVul was the only compared method whose test-pair fix reversal was no longer severe, achieving PairRank of 59.4\%. The other neural detectors evaluated on 564 pairs achieved 37.7\%--45.6\%; see Table~\ref{tab:main-algorithms}. On the 433-pair subset, CFVul, TaCCS-DFA and Vul-LMGNN achieved PairRank of 60.6\%, 37.2\% and 36.6\%, respectively. CFVul's length-matched AUROC was 0.675, comparable to CodeBERT's 0.670. Its overall AUROC was 0.806, below CodeBERT's 0.836. The actual-length output from the same checkpoint achieved overall AUROC of 0.836. The length-standardised and actual-length outputs had length-matched AUROC of 0.675 and 0.674. Their overall AUROC difference arose because the length-standardised output no longer used actual length through the explicit length input.

None of the existing methods constrained the vulnerability-probability ordering of the same function's two versions during training. Standard fine-tuning and structural fusion did not account for function length. SVulD separated the versions in representation space without constraining their vulnerability probabilities. Length-robust training adjusted losses by function-length stratum, comparing different functions. CFVul's content constraint directly targets the vulnerability-probability ordering of the two versions of the same function during training.

\subsubsection{Weakening of the length shortcut}\label{sec:rq2-response}
\begin{figure*}[width=\textwidth,pos=!t]
\centering
\includegraphics[width=0.95\textwidth]{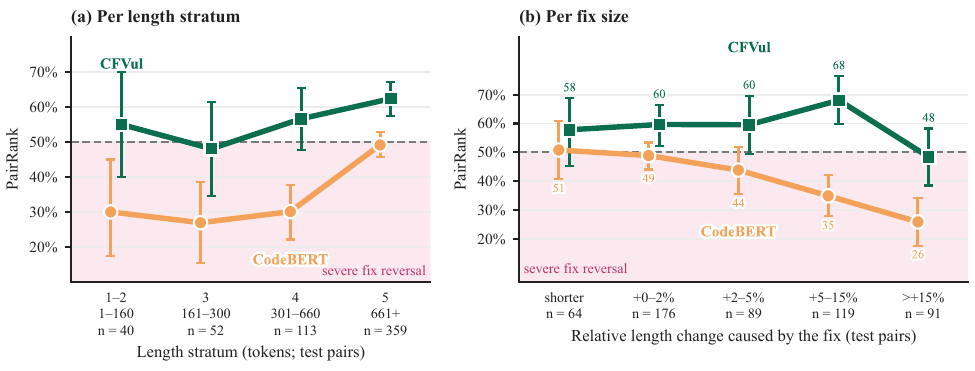}
\caption{Grouped PairRank for CodeBERT and CFVul on test fix pairs; error bars show 95\% bootstrap confidence intervals. (a) Grouping by function-length stratum; stratum 1 contains only 8 pairs and is merged with stratum 2. (b) Grouping by relative length change caused by the fix, excluding 25 equal-length pairs.}
\label{fig:rq2-overview}
\end{figure*}

CFVul weakened the length shortcut, reducing length dependence at both fix-pair and individual-function levels. The value of $\rho_{\Delta}$ fell from CodeBERT's $+0.32$ to CFVul's $-0.05$, while $\rho_{L}$ fell from 0.85 to 0.67; see Table~\ref{tab:main-algorithms}. Across groups of increasing fix-induced relative length change, CodeBERT's PairRank decreased from 51\% to 26\%. CFVul achieved 58\%--68\% in the first four groups without decreasing as fixes grew. In the group exceeding 15\% length increase, CFVul achieved 48\%, compared with CodeBERT's 26\%; see Figure~\ref{fig:rq2-overview}(b). CFVul also had higher PairRank in every function-length stratum; see Figure~\ref{fig:rq2-overview}(a).

\begin{table*}[width=\textwidth,pos=!t]
\centering
\caption{PairRank (\%) for CodeBERT and CFVul on 133 code-transformation pairs. Each group has four columns corresponding to the four fillers.}
\label{tab:rq2-transform}
\small
\setlength{\tabcolsep}{4pt}
\begin{tabular}{@{}lccccccccc@{}}
\toprule
 & Original pair & \multicolumn{4}{c}{Length only ($x^{+}$ vs $\tilde{x}$)} & \multicolumn{4}{c}{Content only ($\tilde{x}$ vs $x^{-}$)} \\
\cmidrule(lr){3-6}\cmidrule(l){7-10}
Detector & $x^{+}$ vs $x^{-}$ & \shortstack{Sampled\\real\\comments} & \shortstack{LLM-\\generated\\comments} & \shortstack{Unused\\variable\\declarations} & \shortstack{Hyphen-\\only\\comments} & \shortstack{Sampled\\real\\comments} & \shortstack{LLM-\\generated\\comments} & \shortstack{Unused\\variable\\declarations} & \shortstack{Hyphen-\\only\\comments} \\
\midrule
CodeBERT & 12 & 2 & 14 & 5 & 15 & 83 & 51 & 71 & 28 \\
CFVul & 50 & 26 & 53 & 22 & 30 & 80 & 42 & 80 & 71 \\
\bottomrule
\end{tabular}
\end{table*}

Code-transformation experiments showed a substantially smaller padding-induced rise in vulnerability probability for CFVul than for CodeBERT. We used the 133 pairs and sampled real comments from Section~\ref{sec:rq1-cause}, separating the original difference into length-only and content-only comparisons. Table~\ref{tab:rq2-transform} reports the results. Original-pair PairRank was 12\% for CodeBERT and 50\% for CFVul. The length-only comparison yielded 2\% and 26\%, respectively; the content-only comparison yielded 83\% and 80\%. The detectors had comparable content contributions, while CFVul's length contribution was substantially weaker. Its original-pair PairRank gain therefore came from the weaker length contribution. CFVul's length-only PairRank remained below 50\%, with padded $\tilde{x}$ receiving higher vulnerability probability in most pairs. The length constraint controls actual length through the explicit length input, but inserted comments can still raise vulnerability probability through the code representation. With the other three fillers, CFVul's length-only PairRank was 53\%, 22\% and 30\%, exceeding CodeBERT's 14\%, 5\% and 15\%. In the content-only comparison, CFVul achieved 42\%, 80\% and 71\%, compared with CodeBERT's 51\%, 71\% and 28\%.

RQ2 shows that CFVul achieves PairRank of 59.4\% on PrimeVul test fix pairs, the highest among the compared methods. Fix reversal on these pairs is no longer severe, and length-matched AUROC is comparable to CodeBERT's. CFVul's $\rho_{\Delta}$ is close to 0, and padding raises its vulnerability probabilities substantially less than CodeBERT's. The PairRank gain comes from weakening the length shortcut.

\subsection{RQ3: Roles of the CFVul components}\label{sec:rq3}\label{sec:rq3-overview}\label{sec:rq3-ablation}

\begin{figure*}[width=\textwidth,pos=!t]
\centering
\includegraphics[width=0.92\textwidth]{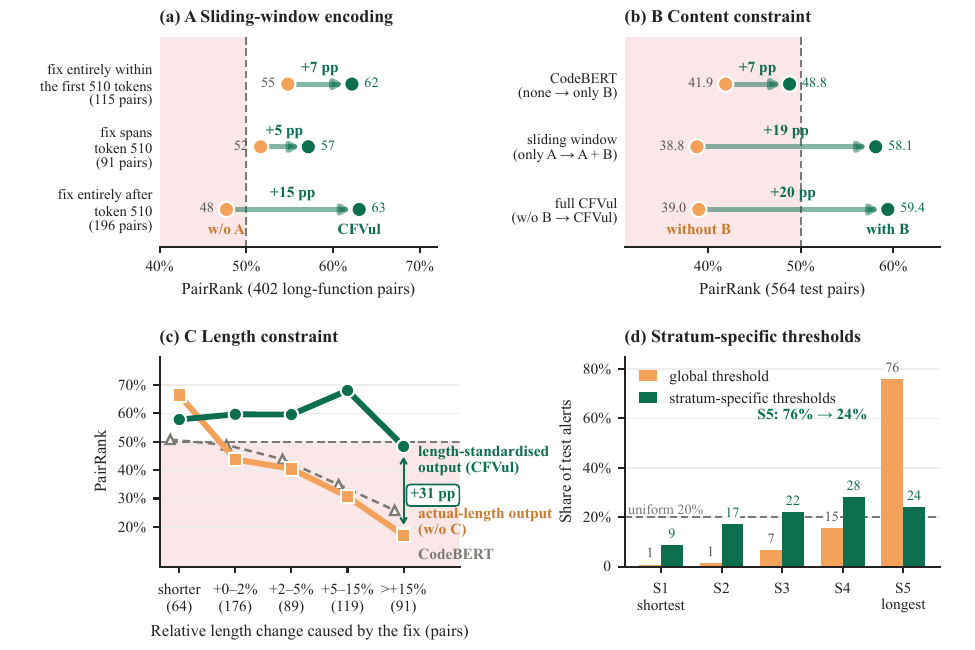}
\caption{Roles of CFVul components and stratum-specific thresholds. (a) Sliding-window encoding A, grouping long-function pairs by fix location relative to the first 510 tokens. (b) Content constraint B, showing PairRank before and after adding B under three settings. (c) Length constraint C, grouped by fix-induced relative length change; grey dashed lines show CodeBERT. (d) Share of test alarms in each function-length stratum at a validation false positive rate of 1\%.}
\label{fig:modules}
\end{figure*}

\begin{table*}[width=\textwidth,pos=!t]
\centering
\caption{Ablation study on the PrimeVul test set. Len.\ input denotes an explicit length input to the classification head. Rows without A use a truncated CodeBERT encoder and omit the window classifier and invariant head.}
\label{tab:main-ablation}
\setlength{\tabcolsep}{2.8pt}
\begin{tabular}{@{}lccccccccccc@{}}
\toprule
 & \multicolumn{4}{c}{Components} & \multicolumn{4}{c}{Length-aware metrics (primary)} & \multicolumn{3}{c}{Conventional metrics} \\
\cmidrule(lr){2-5}\cmidrule(lr){6-9}\cmidrule(l){10-12}
Variant & A & B & \shortstack{Len.\\input} & C & PairRank (\%)$\uparrow$ & \shortstack{Length-matched\\AUROC$\uparrow$} & $\rho_{\Delta}$ & $\rho_{L}$ & Precision & Recall & F1 \\
\midrule
\textbf{All components (ours)} & \checkmark & \checkmark & \checkmark & \checkmark & \textbf{59.4} & \textbf{0.675} & $\mathbf{-0.05}$ & \textbf{0.67} & \textbf{0.180} & \textbf{0.209} & \textbf{0.193} \\
\midrule
\multicolumn{12}{@{}l}{\emph{Remove one component}} \\
w/o A (truncated CodeBERT) & -- & \checkmark & \checkmark & \checkmark & 48.9 & 0.627 & $+0.14$ & 0.84 & 0.170 & 0.275 & 0.210 \\
w/o B (no pair loss) & \checkmark & -- & \checkmark & \checkmark & 39.0 & 0.667 & $+0.29$ & 0.70 & 0.166 & 0.285 & 0.210 \\
w/o C (actual-length output $\sraw$) & \checkmark & \checkmark & \checkmark & -- & 39.4 & 0.674 & $+0.40$ & 0.88 & 0.202 & 0.265 & 0.229 \\
\midrule
\multicolumn{12}{@{}l}{\emph{Keep some components}} \\
A + B (sliding window + pair loss) & \checkmark & \checkmark & -- & -- & 58.1 & 0.632 & $-0.11$ & 0.83 & 0.182 & 0.275 & 0.219 \\
only A (sliding window + BCE) & \checkmark & -- & -- & -- & 38.8 & 0.687 & $+0.35$ & 0.78 & 0.161 & 0.341 & 0.219 \\
only B (CodeBERT + pair loss) & -- & \checkmark & -- & -- & 48.8 & 0.642 & $+0.08$ & 0.84 & 0.206 & 0.201 & 0.203 \\
none (CodeBERT) & -- & -- & -- & -- & 41.9 & 0.670 & $+0.32$ & 0.85 & 0.181 & 0.296 & 0.225 \\
\midrule
\multicolumn{12}{@{}l}{\emph{Remove auxiliary regularisers}} \\
w/o invariant head and window classifier & \checkmark & \checkmark & \checkmark & \checkmark & 54.9 & 0.652 & $-0.02$ & 0.73 & 0.154 & 0.216 & 0.180 \\
\midrule
\multicolumn{12}{@{}l}{\emph{Variants of B}} \\
own lengths, not a common one & \checkmark & \checkmark & \checkmark & \checkmark & 59.0 & 0.674 & $-0.09$ & 0.69 & 0.175 & 0.286 & 0.218 \\
weight 0.05 instead of 0.2 & \checkmark & \checkmark & \checkmark & \checkmark & 44.1 & 0.685 & $+0.17$ & 0.48 & 0.178 & 0.200 & 0.188 \\
weight 0.5 instead of 0.2 & \checkmark & \checkmark & \checkmark & \checkmark & 64.3 & 0.600 & $-0.23$ & 0.61 & 0.121 & 0.213 & 0.154 \\
\bottomrule
\end{tabular}
\end{table*}

We examined the three components through ablation; Table~\ref{tab:main-ablation} and Figure~\ref{fig:modules} report the results. The variants w/o A, w/o B and w/o C remove sliding-window encoding, the content constraint and the length constraint, respectively. Removing any component reduced PairRank to around 50\% or below. Removing both auxiliary regularisers, the invariant head and window classifier, yielded 54.9\%.

Sliding-window encoding exposes fix content beyond the truncated prefix. Removing A reduced PairRank from 59.4\% to 48.9\%. When fixes lay entirely beyond the prefix, w/o A achieved 48\% and CFVul achieved 63\%. This was the largest difference among the three fix-location groups; see Figure~\ref{fig:modules}(a).

The content constraint was the primary source of PairRank improvement. Adding it to CodeBERT increased PairRank from 41.9\% to 48.8\%. Adding it to sliding-window encoding increased PairRank from 38.8\% to 58.1\%. Removing it from the full detector reduced PairRank from 59.4\% to 39.0\%; see Figure~\ref{fig:modules}(b). With sliding-window encoding, the gain from the content constraint increased from 7 to 19 percentage points. Using each version's actual length in the pairwise loss yielded PairRank of 59.0\%, close to the result with common length $\bar L$. The constraint's effect therefore mainly came from paired supervision itself.

The length constraint reduced the dependence of individual-function vulnerability probabilities on length. The w/o C variant and full detector used the same checkpoint. They differed only in whether inference supplied actual or reference lengths to the explicit length input; see Table~\ref{tab:main-ablation}. Actual lengths yielded $\rho_{L}$ of 0.88, $\rho_{\Delta}$ of $+0.40$ and PairRank of 39.4\%. With the length constraint, these became 0.67, $-0.05$ and 59.4\%, respectively. Length-matched AUROC was 0.674 and 0.675 for the two outputs. Across groups of increasing fix-induced relative length change, actual-length PairRank decreased. For fixes increasing length by more than 15\%, it was 31 percentage points below CFVul; see Figure~\ref{fig:modules}(c). The A + B classification head had no explicit length input. Its PairRank was 58.1\%, $\rho_{\Delta}$ was $-0.11$, and $\rho_{L}$ was 0.83, close to CodeBERT's 0.85. Its length-matched AUROC was 0.632. Adding the content constraint to sliding-window encoding reduced $\rho_{\Delta}$ from $+0.35$ to $-0.11$. The explicit length input and length constraint further reduced individual-function $\rho_{L}$ to 0.67 and increased length-matched AUROC to 0.675.

The pairwise-loss weight determines the balance between PairRank and discriminative ability. A weight of 0.05 yielded PairRank of 44.1\%. At 0.5, PairRank reached 64.3\%, while length-matched AUROC fell to 0.600 and F1 to 0.154.

Stratum-specific deployment thresholds changed the distribution of alarms across function-length strata. Alarm thresholds were quantiles of validation-negative vulnerability probabilities, targeting a false positive rate of 1\%. The global threshold used the 99th percentile across all validation negatives. Stratum-specific thresholds used the 99th percentile within each stratum, controlling false positive rates separately. With a global threshold, 76\% of CFVul alarms occurred in stratum 5. Stratum-specific thresholds reduced this to 24\%; see Figure~\ref{fig:modules}(d).

RQ3 clarifies the roles of the three components. Sliding-window encoding covers complete functions and preserves fix information beyond the truncated prefix. Under a common explicit length input, the content constraint encourages higher vulnerability probabilities for vulnerable versions. At inference, the length constraint averages vulnerability probabilities over fixed reference lengths, reducing output dependence on actual length.
\subsection{RQ4: Applying CFVul to other encoders and datasets}

\subsubsection{Application to other encoders and datasets}\label{sec:rq4-encoders}
\begin{table}[pos=!t]
\centering
\caption{Results for five models before and after applying CFVul. Each cell gives the original model / the model with CFVul. Length $+{>}15\%$ denotes the 91 test fix pairs whose fixes increase function length by more than 15\%.}
\label{tab:rq4-encoders}
\scriptsize\setlength{\tabcolsep}{2.2pt}
\begin{tabular}{@{}lcccc@{}}
\toprule
 & \multicolumn{2}{c}{PairRank (\%)$\uparrow$} & & \\
\cmidrule(lr){2-3}
Model & \shortstack{All\\pairs} & \shortstack{Length\\$+{>}15\%$} & $\rho_{\Delta}$ & \shortstack{Length-matched\\AUROC$\uparrow$} \\
\midrule
CodeBERT      & 41.9/59.4 & 25.8/48.4 & $+0.32$/$-0.05$ & 0.670/0.675 \\
UniXcoder     & 42.7/51.2 & 28.0/34.1 & $+0.28$/$+0.08$ & 0.687/0.667 \\
GraphCodeBERT & 45.6/55.9 & 31.9/51.6 & $+0.22$/$-0.01$ & 0.710/0.694 \\
CodeT5        & 42.3/58.4 & 19.2/60.4 & $+0.36$/$-0.13$ & 0.683/0.707 \\
\shortstack[l]{Vul-LMGNN\\(CodeBERT)$^\dagger$} & 36.6/53.7 & 30.0/52.3 & $+0.38$/$-0.05$ & 0.643/0.633 \\
\bottomrule
\end{tabular}
\\[2pt]
{\footnotesize $^\dagger$ Evaluation uses the same subset as Table~\ref{tab:main-algorithms}; Length $+{>}15\%$ contains 65 of these pairs.}
\end{table}

\begin{table*}[width=\textwidth,pos=!t]
\centering
\caption{Results across datasets. Upper section: each cell gives original CodeBERT / CodeBERT with CFVul; Length $+{>}15\%$ follows Table~\ref{tab:rq4-encoders}. Test-pair counts appear in Table~\ref{tab:datasets}. Lower section: adjustments to the BigVul training set, retraining CodeBERT only. The final columns give the adjusted training-set ratio of vulnerability prevalence in the longest to the shortest function-length stratum and Fixed included.}
\label{tab:cross}
\setlength{\tabcolsep}{4pt}
\begin{tabular}{@{}lccccccc@{}}
\toprule
 & \multicolumn{2}{c}{PairRank (\%)$\uparrow$} & $\rho_{\Delta}$ & \shortstack{Length-matched\\AUROC$\uparrow$} & F1 & \shortstack{Vul.-rate\\ratio} & \shortstack{Fixed\\included} \\
\cmidrule(lr){2-3}
Dataset & All pairs & Length $+{>}15\%$ &  &  &  &  &  \\
\midrule
PrimeVul & 41.9\,/\,59.4 & 26\,/\,48 & $+0.32$\,/\,$-0.05$ & 0.670\,/\,0.675 & 0.225\,/\,0.193 & 34.2$\times$ & 78\% \\
BigVul & 60.2\,/\,69.9 & 82\,/\,91 & $-0.40$\,/\,$-0.55$ & 0.890\,/\,0.896 & 0.572\,/\,0.590 & 6.1$\times$ & 98\% \\
ReVeal & 47.4\,/\,58.2 & 28\,/\,35 & $+0.21$\,/\,$+0.23$ & 0.779\,/\,0.776 & 0.329\,/\,0.354 & 7.0$\times$ & 94\% \\
MegaVul & 42.9\,/\,53.7 & 26\,/\,48 & $+0.41$\,/\,$+0.03$ & 0.666\,/\,0.643 & 0.293\,/\,0.175 & 13.4$\times$ & 91\% \\
\midrule
\multicolumn{8}{@{}l}{\emph{Adjustments to the BigVul training set (CodeBERT)}} \\
original training set & 60.2 & 82 & $-0.40$ & 0.890 & 0.572 & 6.1$\times$ & 98\% \\
higher vul.-rate ratio & 52.7 & 61 & $-0.17$ & 0.884 & 0.559 & 32.6$\times$ & 98\% \\
no fixed versions & 47.8 & 51 & $-0.09$ & 0.885 & 0.572 & 6.9$\times$ & 0\% \\
both & 45.3 & 44 & $+0.05$ & 0.886 & 0.571 & 37.2$\times$ & 0\% \\
\bottomrule
\end{tabular}
\end{table*}

CFVul's sliding-window encoding and two constraints do not depend on CodeBERT's internal architecture. We applied the method to three other pretrained sequence models and Vul-LMGNN; see Table~\ref{tab:rq4-encoders}. For sequence models, the corresponding pretrained encoder replaced the shared window encoder, with all other components unchanged. For Vul-LMGNN, sliding-window encoding replaced the sequence branch, retaining the graph branch and original fusion weights. The loss and both constraints were computed on the fusion output. With CFVul's encoding and counterfactual constraints, all five models improved PairRank from 36.6\%--45.6\% to 51.2\%--59.4\%, exceeding 50\%. UniXcoder increased from 42.7\% to 51.2\%. Originally, $\rho_{\Delta}$ ranged from $+0.22$ to $+0.38$. After applying CFVul, it ranged from $-0.13$ to $+0.08$, indicating weaker fix-pair length dependence in all five models. For fixes increasing length by more than 15\%, PairRank increased from 19.2\%--31.9\% to 34.1\%--60.4\%. CodeT5 increased from 19.2\% to 60.4\%. Length-matched AUROC changed by at most 0.03 after applying the method.

PairRank gains were consistent with the original models' length dependence. CodeBERT, CodeT5 and Vul-LMGNN had original $\rho_{\Delta}$ from $+0.32$ to $+0.38$, and gained 16--18 percentage points in PairRank. UniXcoder and GraphCodeBERT had original $\rho_{\Delta}$ of $+0.28$ and $+0.22$, and gained 9 and 10 percentage points, respectively.

CFVul also improved PairRank on the other three datasets; see the upper section of Table~\ref{tab:cross}. On each dataset, we separately trained the original CodeBERT detector and its CFVul counterpart. PairRank increased by 18, 10, 11 and 11 percentage points on PrimeVul, BigVul, ReVeal and MegaVul, respectively. PrimeVul, with the largest ratio of vulnerability prevalence in the longest to the shortest function-length stratum, had the largest gain. On PrimeVul, $\rho_{\Delta}$ fell from $+0.32$ to $-0.05$; on MegaVul, it fell from $+0.41$ to $+0.03$. On BigVul it changed from $-0.40$ to $-0.55$, and on ReVeal from $+0.21$ to $+0.23$. Length-matched AUROC was comparable to CodeBERT's on PrimeVul, BigVul and ReVeal, but decreased from 0.666 to 0.643 on MegaVul. F1 increased on BigVul and ReVeal. On PrimeVul and MegaVul, it decreased from 0.225 and 0.293 to 0.193 and 0.175, respectively. The actual-length output from the same PrimeVul checkpoint had F1 of 0.229. The F1 decrease arose because the length-standardised output no longer used actual length through the explicit length input; see Table~\ref{tab:main-algorithms}.

\subsubsection{Data conditions that expose fix reversal}\label{sec:rq4-datasets}
\begin{figure*}[width=\textwidth,pos=!t]
\centering
\includegraphics[width=0.86\textwidth]{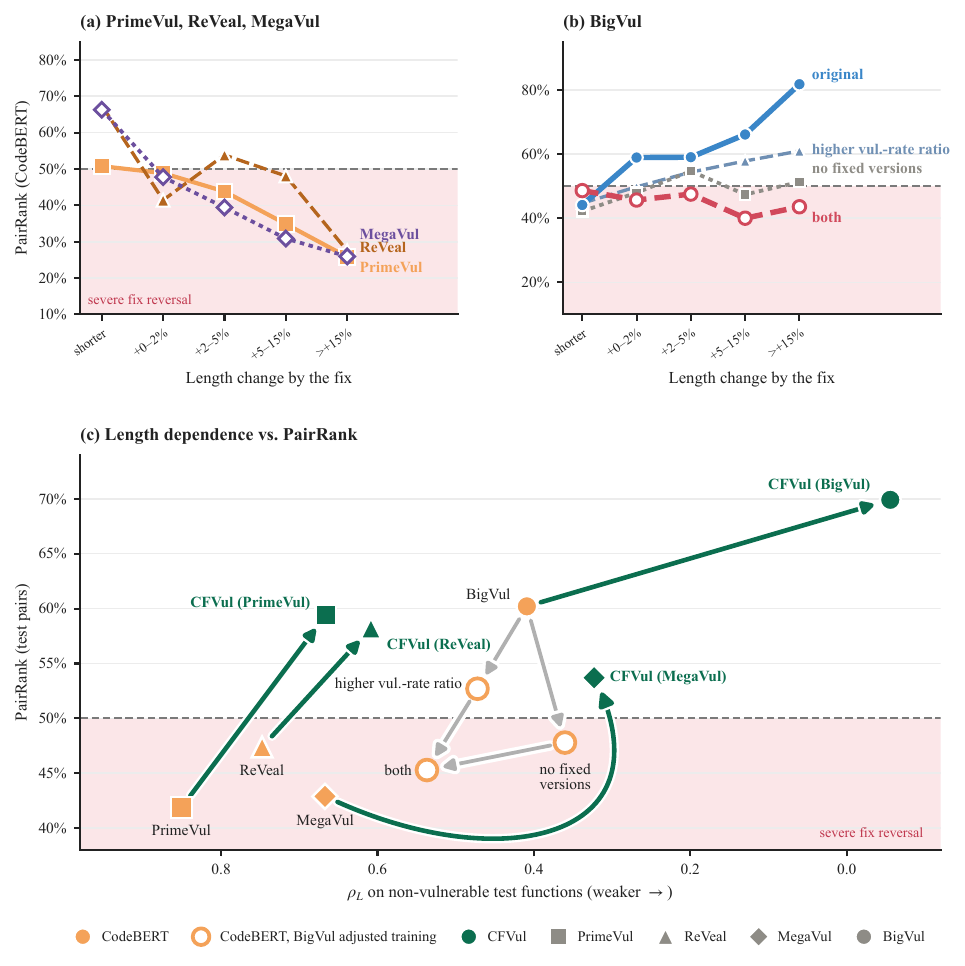}
\caption{Fix reversal across datasets. (a, b) CodeBERT PairRank grouped by relative length change caused by the fix. Panel (b) shows the original BigVul training set and three training-set adjustments. (c) Relationship between $\rho_{L}$ and test-pair PairRank. Grey arrows denote BigVul training-set adjustments; green arrows connect the original CodeBERT detector to its CFVul counterpart on each dataset.}
\label{fig:cross-length}
\end{figure*}

Training-set length--label associations and the retention of fixed-version negatives may affect how clearly fix reversal is exposed. Based on the length-shortcut analysis and dataset observations, we hypothesise two influences. A higher ratio of vulnerability prevalence in the longest to the shortest training-set function-length stratum may make fix reversal more apparent. The number of fixed versions retained as negatives may also affect length dependence. CodeBERT's fix reversal on BigVul was not severe; see Table~\ref{tab:cross} and Figure~\ref{fig:cross-length}(b). This behaviour may reflect its training-data characteristics.

BigVul's ratio of vulnerability prevalence in the longest to the shortest function-length stratum was 6.1, well below PrimeVul's 34.2 and MegaVul's 13.4. Vulnerability prevalence therefore increased more gradually with length. In BigVul, 98\% of vulnerable training functions had fixed versions included as negatives, the highest proportion among the four datasets; see Table~\ref{tab:datasets}. Retaining many longer non-vulnerable functions also reduced fix reversal to some extent. To examine these two training-data characteristics, we adjusted the BigVul training set and retrained CodeBERT. The lower section of Table~\ref{tab:cross} reports the results; Appendix~\ref{app:repro} describes construction details.

After retraining, both adjustments decreased PairRank. Increasing only the ratio of vulnerability prevalence in the longest to the shortest function-length stratum reduced PairRank from 60.2\% to 52.7\%. Removing only fixed-version negatives reduced it to 47.8\%. Applying both adjustments reduced it further to 45.3\%. For fixes increasing length by more than 15\%, PairRank fell from 82\% to 44\%, and $\rho_{\Delta}$ changed from $-0.40$ to $+0.05$. Fix reversal became more apparent. The original training set's trend of increasing PairRank with larger fixes disappeared. All groups with length-increasing fixes had PairRank below 50\%; see Figure~\ref{fig:cross-length}(b). Across the three adjustments, length-matched AUROC ranged from 0.884--0.886 and F1 from 0.559--0.572.

The correlation between vulnerability probability and length explains some changes, although the adjustments had different effects. CodeBERT trained on the three datasets with severe fix reversal had higher $\rho_L$ than CodeBERT trained on BigVul; see Figure~\ref{fig:cross-length}(c). On BigVul, increasing the prevalence ratio or applying both adjustments increased $\rho_L$ and decreased PairRank. Removing fixed-version negatives decreased PairRank without increasing $\rho_L$. These training-set adjustment results support the proposed hypothesis.

RQ4 shows that CFVul improves PairRank across five models and four datasets. All five models exceeded 50\% PairRank. CFVul also exceeded 50\% on test fix pairs in all four datasets. Fix reversal was no longer severe on PrimeVul, ReVeal and MegaVul. BigVul training-set adjustments showed that increasing the ratio of vulnerability prevalence in the longest to the shortest function-length stratum and removing fixed-version negatives made fix reversal more apparent.
\section{Discussion}\label{sec:discussion}
Fix pairs test a detector's assessment of vulnerability-status changes in the same function before and after fixing. In PrimeVul, function length correlates with vulnerability labels. Length alone achieves overall AUROC close to a fine-tuned model; see Table~\ref{tab:main-algorithms}. Fixes usually lengthen functions, so length dependence conflicts with the change in vulnerability status after fixing. A detector can therefore combine high conventional classification metrics with poor pairwise ordering.

CFVul's ranking improvement mainly comes from reduced length effects. With sampled real comments, CFVul and CodeBERT perform comparably in the content-only comparison, while CFVul responds substantially less to padding; see Table~\ref{tab:rq2-transform}. Length standardisation blocks actual length from influencing the output through the explicit length input. Padding can still change the output through the code representation. Evaluation should combine length padding and equal-length comparisons to test separately whether length effects weaken and content discrimination improves.

\paragraph{Threats to validity}
Regarding internal and statistical conclusion validity, all models were reproduced under a common protocol, so metrics may differ from those reported for the original methods. Main results come from a single training run; paired bootstrap confidence intervals exclude training randomness. Regarding construct validity, auxiliary window labels are derived from line-level diffs, while fix pairs retain dataset labels. Unrelated changes and labelling errors may affect supervision and the interpretation of pairwise ordering. We did not compare effective fixes with fix-like changes that retain the vulnerability. PairRank improvements therefore do not assess detectors' ability to judge fix validity.

Regarding external validity, the experiments are limited to C/C++ function-level binary classification. Missing calling context can restrict vulnerability assessment~\citep{risse2025top}. Project overlap between training and test sets may inflate evaluation results relative to performance on unseen projects. Separate training on each dataset does not establish direct applicability of the same model to new projects or datasets. Code transformations cover only pairs with longer fixed versions, all changes within the input prefix, and transformed versions within the input-length limit. This selection limits the scope of the analysis of fix-reversal causes.

\section{Conclusion}\label{sec:conclusion}
We examined function-level vulnerability detectors using fix pairs. Fixed versions often received higher vulnerability probabilities than their vulnerable counterparts, with more severe reversal for larger fix-induced length increases. Code-transformation experiments indicate that the length shortcut is the main source of fix reversal in the examined pairs. During training, CFVul applies a content constraint to distinguish vulnerable and fixed versions under a common explicit length input. During inference, its length constraint weakens the influence of actual length on the output. On PrimeVul, CFVul achieved PairRank of 59.4\%, the highest among the compared methods. Length-matched AUROC remained comparable to CodeBERT's, indicating improved ordering through weaker length effects while retaining content discrimination. For fixes increasing length by more than 15\%, PairRank rose from 25.8\% to 48.4\%. PairRank also improved with four additional models and three additional datasets.

Future work will further reduce residual length dependence in code representations. It will compare effective fixes with fix-like changes retaining vulnerabilities and extend fix-pair evaluation to other programming languages.

\FloatBarrier

\section*{Declaration of competing interest}
The authors declare that they have no known competing financial interests or personal relationships that could have appeared to influence the work reported in this paper.

\appendix
\counterwithin{table}{section}
\counterwithin{figure}{section}

\section{Reproduction details}\label{app:repro}
We used the complete official PrimeVul splits without filtering or sampling (184,427 training, 25,430 validation and 25,911 test functions). Joern CPG generation succeeded on 175,786 training, 23,948 validation and 24,788 test functions, yielding 3,770, 480 and 433 fix pairs with CPGs for both versions. For BigVul (188,636 original rows), we split $8:1:1$ by \texttt{commit\_id}, stripped trailing spaces per line and at function ends (in original BigVul, 59\% of functions changed by fixing commits had trailing spaces, compared with 2.3\% of unchanged functions), skipped 250 pairs that were identical after whitespace compression, and retained 199,286 functions without deduplication. For ReVeal (2,240 vulnerable and 20,494 non-vulnerable pre-tokenised functions), whitespace was left unchanged; same-label duplicates were collapsed and 351 label-conflicting code strings were dropped in both copies, leaving 22,032 functions split $8:1:1$ by patch \texttt{hash} and paired within \texttt{hash} and function name (1,369, 197 and 195 pairs, using each function at most once). For MegaVul (353,873 records from the 2024-04 C/C++ release), we split $8:1:1$ by \texttt{commit\_hash}, stripped trailing spaces per line (8\%--9\% of vulnerable or fixed functions versus 2\% of unchanged functions), removed 1,773 negatives identical to vulnerable functions, deduplicated 14,471 remaining copies in the order vulnerable, fixed and unchanged, and excluded 352 identical pairs plus 1,643 pairs with one removed member, leaving 355,604 functions. Evaluation used 564 PrimeVul test pairs (433 in the CPG subset), 133 code-transformation pairs (74 validation and 59 test from 445 candidates), and 1,222, 195 and 1,712 test pairs on BigVul, ReVeal and MegaVul, respectively.

\begin{table}[pos=!hbp]
\centering
\caption{Hyperparameter settings for CFVul.}
\label{tab:hparams}
\scriptsize\setlength{\tabcolsep}{3pt}
\begin{tabular}{@{}lp{0.58\linewidth}c@{}}
\toprule
Symbol & Meaning & Value \\
\midrule
$W$ & maximum window length in body tokens & 128 \\
-- & maximum window stride $W/2$ & 64 \\
$h$ & attention hidden size & 128 \\
$\lambda_{\mathrm{inv}}$ & invariant-head risk weight & 1 \\
-- & length environments of the invariant head & 4 \\
$\lambda_{\mathrm{pair}}$ & pair loss weight & 0.2 \\
$\lambda_{\mathrm{aux}}$ & window classifier loss weight & 0.05 \\
$r_{\mathrm{ins}}$ & lines on each side of a pure insertion marked as changed & 1 \\
$\Kref$ & number of reference lengths & 32 \\
ref & reference-length distribution & training empirical \\
$\delta$ & log-length matching tolerance of $R_k$ & 0.05 \\
\bottomrule
\end{tabular}
\end{table}

On PrimeVul, every neural detector used the official splits and was trained for 3 full epochs with AdamW (weight decay $0.01$, gradient clipping $1.0$, and $6\%$ linear warmup where applicable). The effective batch size was 64 functions. We selected the checkpoint with the highest validation mean length-matched AUROC across the five strata; other hyperparameters followed the original methods. The learning rate was $2\times10^{-5}$ for the sequence baselines; in cross-dataset CFVul runs, the encoder used $2\times10^{-5}$, the pooling, head and invariant-direction parameters used $10^{-4}$, and the invariant intercept used $10^{-3}$. Vul-LMGNN used $2\times10^{-5}$ for CodeBERT, $2.1\times10^{-5}$ for the classifier head and $10^{-4}$ (weight decay $1.3\times10^{-6}$) for its graph branch; TaCCS-DFA retained the locked encoder and head learning rates ($2\times10^{-5}$ and $2.1\times10^{-5}$).
For PrimeVul's 184,427 training functions, groups of 64 functions with the final group retained imply 2,882 optimisation groups per epoch, or 8,646 planned updates over 3 epochs (Vul-LMGNN on 175,786 CPG functions with incomplete batches dropped ran 2,746 updates per epoch, 8,238 total). Under the same 3-epoch, batch-64 scheme, planned updates were 7,293 on BigVul (155,575 training functions), 789 on ReVeal (16,771) and 13,350 on MegaVul (284,762).
Each model's main metrics came from one training run, using seed 13. Experiments ran on one NVIDIA RTX 3090 with 24\,GB memory.
For CodeT5, we used its encoder and averaged token representations to obtain function or window representations. Table~\ref{tab:hparams} lists CFVul's hyperparameters. Inference reference lengths were sampled once from the training-set empirical distribution.

CFVul training uses gradient caching. Each optimisation group contains 64 functions, keeping both versions of each fix pair together. After encoding all windows and caching their representations, we compute the group loss and its gradients with respect to window representations. We then re-encode each window and propagate the cached gradients through the encoder. Parameters are updated once after all gradients for the group have been computed.
At inference, each window is encoded once. The same function-content representation is evaluated at its actual length and the fixed reference lengths to obtain the two outputs.

\paragraph*{Adjusting the BigVul training set}
Fixed versions in BigVul were added from post-fix functions in the original data and labelled non-vulnerable. During construction, their sample identifiers were prefixed with \texttt{f}, followed by the original record number of the corresponding vulnerable version. To remove fixed versions alone, we deleted training records with this marker while retaining all vulnerable functions and other non-vulnerable functions. Fixed included uses all vulnerable training functions as its denominator. Its numerator counts those whose corresponding fixed version also occurs in the training set. Removing fixed versions reduced this proportion to 0\%, leaving no complete fix pairs in training. Retained vulnerable functions and other non-vulnerable functions continued to support classification training.

To increase the ratio of vulnerability prevalence in the longest to the shortest function-length stratum, we targeted the relative prevalence pattern across PrimeVul's function-length strata. All vulnerable functions in the longest stratum were retained; vulnerable functions in other strata were retained probabilistically. Fixed versions corresponding to removed vulnerable functions were also removed. The combined adjustment applied both vulnerable-function downsampling and fixed-version removal. After both operations, sampled vulnerable functions and other non-vulnerable functions were retained. Every setting retrained CodeBERT using only function-level labels. Validation and test functions and their fix pairs remained unchanged; PairRank was computed on the original test fix pairs.

\FloatBarrier
\section{Additional results for RQ1}\label{app:rq1}
\begin{figure*}[width=\textwidth,pos=!htbp]
\centering
\includegraphics[width=0.92\textwidth]{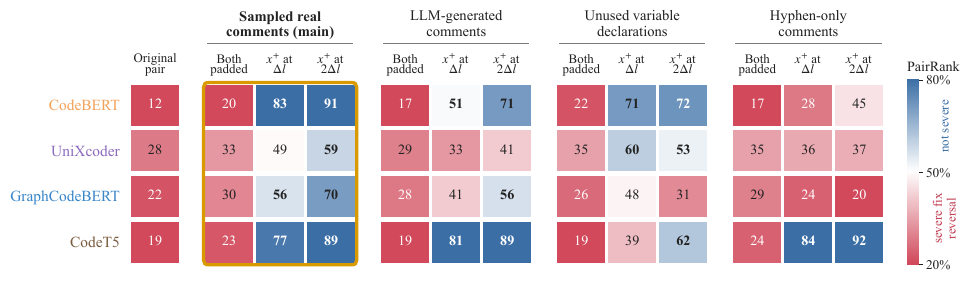}
\caption{PairRank for 133 code-transformation pairs under four fillers. Both padded denotes the control with identical padding added to both versions. The labels $x^{+}$ at $\Delta l$ and $x^{+}$ at $2\Delta l$ add $\Delta l$ and $2\Delta l$ tokens only to $x^{+}$. The yellow frame marks the sampled real comments used in the main text.}
\label{fig:iv-fillers}
\end{figure*}

Table~\ref{tab:rq1-dose} gives the group values in Figure~\ref{fig:rq1-dose}. Figure~\ref{fig:iv-fillers} reports the four fillers for the code-transformation experiments in Section~\ref{sec:rq1-cause}. The label $x^{+}$ at $\Delta l$ denotes the equal-length comparison of $\tilde{x}$ against $x^{-}$. LLM-generated comments came from Qwen3-4B-Instruct-2507 using greedy decoding with 8 lines of context on each side of the insertion point. The control with identical padding added to both versions used the same content at their fix locations, preserving the length difference. This ensured that both versions received the inserted content. This separates length changes from the inserted content's own effect. With sampled real comments, control PairRank remained at 20\%--33\%, within the severe fix-reversal region. The increase from padding only $x^{+}$ therefore came from the changed length difference. With the other three fillers, CodeBERT and CodeT5 PairRank increased with padding amount. GraphCodeBERT and UniXcoder also increased with LLM-generated comments. For unused variable declarations, increasing padding from $\Delta l$ to $2\Delta l$ reduced their PairRank from 48\% and 60\% to 31\% and 53\%, respectively. GraphCodeBERT's PairRank with hyphen-only comments was 20\%--24\%, close to its original-pair value of 22\%.

\begin{table*}[width=\textwidth,pos=!htbp]
\centering
\caption{PairRank (\%) for the groups in Figure~\ref{fig:rq1-dose}. Square brackets give 95\% bootstrap confidence intervals. The 25 equal-length pairs are excluded from grouping.}
\label{tab:rq1-dose}
\small
\setlength{\tabcolsep}{4pt}
\begin{tabular}{@{}lccccc@{}}
\toprule
Length change & Pairs & CodeBERT & UniXcoder & GraphCodeBERT & CodeT5 \\
\midrule
All pairs & 564 & 41.9 [39, 45] & 42.7 [39, 46] & 45.6 [42, 49] & 42.3 [39, 46] \\
\midrule
Fixed shorter & 64 & 50.8 [41, 61] & 44.5 [34, 55] & 47.7 [38, 58] & 60.2 [50, 70] \\
$(0, 2\%]$ & 176 & 48.9 [44, 54] & 53.7 [49, 59] & 56.0 [51, 61] & 51.4 [46, 56] \\
$(2\%, 5\%]$ & 89 & 43.8 [36, 52] & 39.3 [30, 48] & 49.4 [41, 58] & 41.0 [33, 50] \\
$(5\%, 15\%]$ & 119 & 34.9 [28, 42] & 34.5 [27, 42] & 38.2 [31, 46] & 36.6 [29, 45] \\
$>15\%$ & 91 & 25.8 [18, 35] & 28.0 [19, 37] & 31.9 [23, 41] & 19.2 [12, 27] \\
\bottomrule
\end{tabular}
\end{table*}

\FloatBarrier

\renewcommand{\bibfont}{\fontsize{8pt}{9.8pt}\selectfont}
\bibliographystyle{elsarticle-num}
\bibliography{refs}

\end{document}